\documentclass[epj]{svjour}
\usepackage{graphicx,amssymb}
\begin{document}
\title{Energy landscapes, lowest gaps, and susceptibility
of elastic manifolds at zero temperature}

\author{E.\ T.\ Sepp\"al\"a and M.\ J.\ Alava}

\institute{Helsinki University of Technology, Laboratory of
Physics, P.O.Box 1100, FIN-02015 HUT, Finland}
\date{Received: December 12, 2000 / Revised version: April 11, 2001}

\abstract{
We study the effect of an external field on $(1+1)$ and $(2+1)$
dimensional elastic manifolds, at zero temperature and with random
bond disorder.  Due to the glassy energy landscape the configuration
of a manifold changes often in abrupt, ``first order'' -type of large
jumps when the field is applied.  First the scaling behavior of the
energy gap between the global energy minimum and the next lowest
minimum of the manifold is considered, by employing exact ground state
calculations and an extreme statistics argument. The scaling has a
logarithmic prefactor originating from the number of the minima in the
landscape, and reads $\Delta E_1 \sim L^\theta [\ln(L_z
L^{-\zeta})]^{-1/2}$, where $\zeta$ is the roughness exponent and
$\theta$ is the energy fluctuation exponent of the manifold, $L$ is
the linear size of the manifold, and $L_z$ is the system height. The
gap scaling is extended to the case of a finite external field and
yields for the susceptibility of the manifolds $\chi_{tot} \sim
L^{2D+1-\theta} [(1-\zeta)\ln(L)]^{1/2}$.  We also present a mean
field argument for the finite size scaling of the {\it first jump
field}, $h_1 \sim L^{d-\theta}$. The implications to wetting in random
systems, to finite-temperature behavior and the relation to
Kardar-Parisi-Zhang non-equilibrium surface growth are discussed.
} 

\authorrunning{E.\ T.\ Sepp\"al\"a and M.\ J.\ Alava}
\titlerunning{Energy landscapes, lowest gaps, and susceptibility of elastic
manifolds at $T=0$}
\maketitle

\section{Introduction}\label{intro}

In this paper we study zero-temperature or ground state elastic
manifolds that are roughened by bulk disorder, in the presence of an
external field. Such objects are relevant in many contexts of
condensed matter and statistical physics~\cite{Fis86,Emig99}. The
essential point here is a competition between the elasticity, which
prefers flat manifolds, and the disorder, which induces wandering in
order to take advantage of the low energy regions in the system.  This
leads to a glassy, complicated energy landscape and the fact that the
quenched randomness dominates thermal effects at low temperatures.  In
two embedding dimensions (2D) such manifolds are under the name
directed polymers (DP)~\cite{KPZ,HaH95,Las98} particularly interesting
through their connection to the celebrated Kardar-Parisi-Zhang (KPZ)
equation of nonlinear surface growth.  Directed polymers have an
experimental realization as vortex-lines in granular
superconductors~\cite{blatter}.  In higher dimensions elastic
manifolds may be best considered as domain walls (DW) in ferromagnets
with quenched impurities. Elastic manifolds have also other
connections to charged density waves, the sine-Gordon model with
disorder, random substrate problems, and vortex lattices, to name but
a few~\cite{blatter,Cardy82,Toner90}.
	
Let us start by introducing the classical spin-half random Ising Hamiltonian
\begin{equation}
{\mathcal H} = - \sum_{\langle ij \rangle} J_{ij} S_i S_j 
- \sum_i H_i S_i,
\label{RHamilton}
\end{equation}
where $J_{ij}$ is the coupling constant between the nearest-neighbor
spins $S_i$ and $S_j$, and $H_i$ is a field assigned to each spin. To
this system we apply antiperiodic or domain wall -enforcing boundary
conditions in one direction. The spins in the opposite boundaries, let
us define in $z$-direction, $z=0$ and $z=L_z$ are forced to be up and
down, respectively.  In the case of ferromagnetic random bond (RB)
Ising systems one has $J_{ij} \geq0$ and $H_i=0$. In the minimum
energy state the spins prefer to align on each side of the induced
domain wall.  When $J_{ij} \lessgtr 0$ the spins become frustrated and
the task to find the ground state (GS) structure is most often related
to spin glass physics~\cite{review,droplet,replica}. 
On the other hand, when for simplicity $J_{ij} = {\rm
const.} = J > 0$, and $H_i \lessgtr 0$ one arrives at random field
(RF) Ising systems. The random field Ising model (RFIM) has an
experimental realization as a diluted antiferromagnet in a field. In
the RFIM the ferromagnetic couplings compete with the random field
contribution, which prefers in the ground state to have the spins to
be oriented towards the field assigned to them.

Here we concentrate mainly on the random bond Ising Hamiltonian, i.e.,
$J_{ij}>0$ and $H=0$, and in some special cases extend the discussion
to RF domain walls as well.  In the simplest case the spins are
located in a square lattice, in $d=2$, or a cube in $d=3$, so that the
lattice orientation is in the \{10\} and \{100\} directions,
respectively.  The elastic manifold is the interface, with the
dimension $D=d-1$, which divides the system in two parts of up and
down spins. At $T=0$ the problem of finding the ground state domain
wall, which minimizes the path consisting of unsatisfied bonds between
the spins on opposite sides of the domain wall becomes ``global
optimization''.  In our case the displacement field is
one-dimensional, $n=1$, but one can certainly think about
generalizations so that the total dimension of a system $d=(D+n)$,
where $n \geq 1$ is the dimension in which the manifold is able to
fluctuate.

The continuum Hamiltonian of elastic manifolds with an external field
and $n=1$ may be written as
\begin{equation}
{\mathcal H} = \int \left[ \frac{\Gamma}{2} \{ \nabla 
z({\bf x}) \}^2 + V_r \{ {\bf x},z({\bf x})\} + h\{z({\bf x})\} 
\right] \, {\rm d}^D{\bf x}. \label{H}
\end{equation} 
The elastic energy is proportional to the area of the interface given
by the first term, and $\Gamma$ is the surface stiffness of the
interface.  The second term of the integrand comes from the random
potential, and the last term accounts for the potential caused by the
external field. The use of random bond disorder means that the random
potential is delta-point correlated, i.e., $\langle V_r ({\bf x},z)V_r
({\bf x'},z') \rangle =2 {\mathcal D} \delta ({\bf x}- {\bf x'})\delta
(z-z')$. In the random field case $\langle V_r ({\bf x},z)V_r ({\bf
x'},z') \rangle \sim \delta ({\bf x}- {\bf x'}) (z-z')$. The
Hamiltonian (\ref{H}) is also applicable to wetting in a three phase
system, where two of the phases are separated by an interface in a
random bulk~\cite{Lip86,Wuttke91,Dole91}. In that case $h(z)$ is
equivalent to the chemical potential, which tries to bind the
interface to the wall, and competes with the random potential, in the
presence of which the interface tends to wander in the low energy
regions of the system.

Below the upper critical dimension the geometric behavior of elastic
manifolds is characterized by the spatial fluctuations. For the
mean-square fluctuations one has
\begin{equation}
w^2 = \left \langle \left[z({\bf x}) - \overline{z({\bf x})} \right ]^2 
\right \rangle \sim L^{2 \zeta},
\label{w}
\end{equation}
where $z({\bf x})$ is the height of the interface with the mean
$\overline{z}$, ${\bf x}$ is the $D=d-n$ dimensional internal
coordinate of the manifold, $L$ is the linear size of the system, and
$\zeta$ is the corresponding roughness exponent. At low temperatures
in $(D+n)=(1+1)$ dimensions with RB disorder, i.e., when one actually
considers a directed polymer in random media~\cite{KPZ,HaH95}, the
roughness exponent is calculated exactly via the KPZ formalism to be
$\zeta=2/3$.  In higher manifold dimensions $D$ with $n=1$ the
functional renormalization group (FRG) calculations give the
approximatively values $\zeta \simeq 0.208(4-D)$ for RB disorder and
$\zeta = (4-D)/3$ for RF disorder~\cite{Fis86,new}. The expression for
$\zeta$ tells also that the upper critical dimension for the elastic
manifold is $D_u=4$.  For manifolds with varying $n$ and $D$ Balents
and D.\ Fisher have derived using FRG $\zeta \simeq
[(4-D)/(n+4)]\{1+(1/4e)2^{-[(n+2)/2]}[(n+2)^2/(n+4)]
[1-\ldots]\}$~\cite{Balents}.  At zero temperature the total average
energy $\overline{E}$ of an elastic manifold equals its free energy
and grows linearly with the system size $L^D$ and its fluctuations
scale for all $n$ as $\Delta E = \left \langle ( E- \overline{E} )^2
\right \rangle^{1/2} \sim L^\theta$, where \cite{HuHe} 
\begin{equation}
\theta = 2 \zeta +D -2
\label{HuHe}
\end{equation}
is the first non-analytic correction to the energy. The same
hyperscaling law holds for RF manifolds, too, in $D>1$
\cite{Seppala98}. Having a positive $\theta$ implies that the
temperature is an irrelevant variable in the renormalization group
(RG) sense and the $T=0$ fixed point dominates. For $D=1$, $n > 2$
there exists a $T_c$, and $T=0$ fixed point dominates only below
$T_c$, likewise always for $n\leq 2$~\cite{Fish91}. At the randomness
dominated pinned phase the temperature is ``dangerously'' irrelevant,
which means that in RG calculations the interesting correlation
functions cannot be obtained by setting $T$ to zero.  Above $T_c$ the
fluctuations become random walk -like with $\zeta=1/2$ and $\theta
=0$. For $D=1$ with $n>1$ there is no exact result existing for the
roughness exponent below $T_c$, and hence whether $\zeta \to 1/2$ and
$\theta \to 0$ for a finite $n_c$ is still an open question -- that
is, what is the upper critical dimension of the KPZ equation.

Elastic manifolds self-average in the sense that the intensive
fluctuations of the roughness and the energy decrease with system
size. However, in this paper we will show that, due to the glassiness
of the energy landscape, for example the behavior of the mean position
of a typical example of a manifold $\overline{z({\bf x})}$ does not
coincide with the disorder average, here denoted by $\langle \,
\rangle$, over many realizations with different random configurations.
Introducing the external field to a random system induces often
drastic changes and one experimental possibility is to study the
ground state behavior e.g. by measuring the susceptibility. In
disordered superconductors the external field is due to the current
density $j$, which drives the vortex lines.

Here we study the elastic manifolds in an external field while being
especially interested in the information it gives about the energy
landscape of the random system~\cite{Hou1}, which is closely related
to the susceptibility.  The external field is applied to the coupling
constants [see Eq. (\ref{H})] so that $J_\perp(z) = J_{random} +
h(z)$, where $J_\perp$ are ones in the $z$-direction and $h$ is the
amplitude of $h(z)$, the strength of the external field. Since we have
the field potential linear in height, $h(z)= hz$, in ferromagnetic
random systems with a domain wall the external field $h(z)$ may be
transformed to a constant external field term $-H\sum_i S_i$ in the
Hamiltonian (\ref{RHamilton}).  Our results are a systematic extension
of early numerical work by Mez\'ard and relate to more recent
discussion of the energy landscapes of directed polymers by Hwa and
D.\ Fisher~\cite{Mez90,Hwa94}.  Due to the glassiness of the energy
landscape the position changes in ``first order type'' large jumps, at
sample-dependent values of the external field.  The second scenario in
which the perturbations take place locally via ``droplet''-like
excitations is ruled out by a scaling argument and numerical results.

We start by studying a specific case where the number of valleys is
fixed to a constant.  We first consider the case without the external
field, and find the scaling behavior of the energy difference of the
global energy minimum and the next lowest minimum of a manifold
numerically and derive it also from an extreme statistics
argument. The scaling of the gap does not only depend on the energy
fluctuation scale, but has in addition a logarithmic factor dependent
on the number of the low energy minima. The gap scaling is extended to
the finite external field case, to derive the susceptibility of the
manifolds under the assumption that the zero-field energy landscape is
still relevant.  The corresponding numerics allows us also to deduce
the effective gap probability distribution without any a priori
assumptions.  We then proceed by constructing a mean field argument
for the finite size scaling of the first jump field in a more general
case.  This agrees with ``grand-canonical'' numerics in which the
manifold is allowed to have an arbitrary ground state position in the
system: the number of valleys in the energy landscape fluctuates and
the important physics is caught by the simple scaling considerations.
Finally, we discuss wetting in random systems in the light of our
results, and compare with the necklace theory of M.\ Fisher and
Lipowsky that applies in the limit of large fields. We also consider
finite temperatures, and relate the physics at finite external fields
to the Kardar-Parisi-Zhang growth problem through the arrival time
mapping. The implications of our results in that case concern the {\em
two first} arrival times and their difference of an interface to a
prefixed height. Note that we have published short accounts
of some of the work contained here earlier\cite{Seppala00,unpublished}.

The paper is organized so that it starts with a short review of the
thermodynamic behavior of the interfaces in random media at zero
temperature in the presence of an applied field, and of the folklore
concerning energy landscapes, in Section~\ref{thermo}.
Section~\ref{glassy} introduces the mean-field level behavior of the
interface when the external field is applied.  The analytical
probability distribution of the first {\it jump} field is derived and
its relation to the susceptibility is discussed in
Section~\ref{argumentti}. In the section also the lowest energy gaps
between the local energy minima and the global minimum are derived
from extreme statistics arguments. The numerical method of
calculating the first jump field $h_1$ of an interface in a fixed
height is introduced in Section~\ref{numerics}. Section~\ref{results}
contains the numerical results of the first jump field in $(1+1)$ and
$(2+1)$ dimensions with the corresponding jump geometry statistics as
well as the energy gaps of the first lowest energy minima.  The
results are compared with the analytical arguments presented in the
previous section. In Section~\ref{first_jump} a mean field argument
for the first jump field of an interface which lies originally in an
arbitrary height is derived and compared with the numerical data and
the evolution of the jump size distributions is studied. As an
application for the finite field case the wetting phenomenon is
considered and the corresponding wetting exponents are numerically
studied in Section~\ref{wetting}. The arguments presented in the paper
are discussed from the viewpoint of interface behavior
at low but finite temperatures in Section~\ref{creep} and in the
context of first
arrival times in KPZ growth in Section~\ref{secKPZ}. The paper is
finished with conclusions in Section~\ref{concl}.

\section{Thermodynamic considerations}\label{thermo}

The interesting behavior arises since in the low temperature phase
e.g.~directed polymers are found to have {\it anomalous} fluctuations
resulting from the regions of the random potential with almost
degenerate energy minima, which are separated by large energy
barriers~\cite{Fish91,Mez90,Hwa94}.  These spatially large-scale
low-energy excitations are rare, but are expected to dominate the
thermodynamic properties and cause large variations in the structural
properties at low temperatures.  We will use the arguments in the next
section in order to study the movement of the manifolds in equilibrium
when an external field is applied.  Hwa and D.\ Fisher~\cite{Hwa94}
derived that for a polymer fixed at one end a small energy excitation
with a transverse scale $\Delta$ ($\simeq l^\zeta$, where $l$ is the
linear length of the excitation), scales as $\Delta^{\theta/\zeta}$
and gives rise to large sample-to-sample variations in the two-point
correlation function and dominates the disorder averages.  The
probability distribution for the sizes of the excitations was found to
be a power-law from the normalization of ``density of states'' of
small energy excitations $W \sim \Delta^{-n} l^{-\theta} \sim
1/\Delta^{n+2-1/\zeta}$, since $\theta =2\zeta -1$, Eq.~(\ref{HuHe}).
This is in contrast to the high temperature phase one, where $W$ is
Gaussian. For the finite field case Hwa and D.\ Fisher argued that the
power-law distribution for the excitations is due to the statistical
tilt symmetry, which means that the random part of the new potential
of the polymer is statistically the same as the old one, when the
polymer is excited by an applied field.  In Ref.~\cite{Mez90} M\'ezard
showed numerically that the energy difference of two copies of the
polymer in the same realization of the random potential scales as
$L^\theta$, where $\theta=1/3$ in $d=2$.  He also showed that the
ground state configuration of a polymer, which is fixed at one end and
applied with an external field $h$ to the end point of the polymer,
i.e., $h\{z({\bf x})\} = hz\delta(x-L)$ in Hamiltonian (\ref{H}),
changes abruptly with a distance $L^\zeta, \zeta=2/3$, when the field
is increased by $L^{\theta}$.  Calculating the variance of the end
point of the polymer $var(z) \sim \langle z^2 \rangle - \langle z
\rangle^2$ M\'ezard found that typically it is zero, i.e., the
ground state configuration does not change.  With the probability
$L^{-\theta}$, there is a sample for which $h=0$ is the critical value
of $h$, and the configuration changes with a large difference compared
to the original state $L^{2\zeta} \sim L^{1+\theta}$,
Eq.~(\ref{HuHe}). Hence the total average variance becomes
$var(z)_{ave} \sim L$. In this paper we generalize such studies as
discussed later. 

The physics of DP's have been shown by Parisi to obey weakly broken
replica symmetry~\cite{Parisi90}, a ``baby-spin glass''
phenomenon. This means that the replica symmetric solution of a DP is
degenerate with the solution with the broken replica symmetry. This is
also evidence for an energy landscape with several, far away from each
other, nearly degenerate local minima.  M\'ezard and Bouchaud studied
later the connection between extreme statistics and one step replica
symmetry breaking in the random energy model (REM)~\cite{derrida}, in
a finite, one-dimensional form \cite{boumez}. They computed using
extreme statistics the probability distribution of the minimum of all
the energies $\frac{\gamma}{2}x^2 + E(x)$, where $E(x)$ are random and
follow a suitable probability distribution so that its tails decay
faster than any power-law, e.g. Gaussian.  This can be seen as a toy
model for an interface in a random media, in which case the quadratic
part plays the role of the elasticity. The minimum of the total energy
is easily seen to be distributed according to the Gumbel
distribution~\cite{Gumbel} of extreme statistics, and likewise for
the position of the interface the shape of the probability
distribution to be approximately Gaussian. In Section~\ref{argumentti}
we will use extreme statistics to study the scaling of the lowest
minima and energy gaps in elastic manifolds in a situation analogous
to that of Bouchaud and M\'ezard.

The response of the manifolds pinned by quenched impurities to
perturbations was discussed in terms of thermodynamic functions by
Shapir~\cite{Shapir91}, including the susceptibility of the
manifolds. The formal definition of the susceptibility for a $D$
dimensional manifold in a $d$ dimensional embedding space reads:
\begin{equation}
\chi = \lim_{h \to 0+} \left \langle \frac{\partial m}{\partial h} 
\right \rangle,
\label{Eqsuskis}
\end{equation} 
where the change in the magnetization of the whole $d$ dimensional
system is calculated in the limit of the vanishing external field from
the positive side and the brackets imply disorder average. In
Section~\ref{results} we derive the susceptibility based on energy
landscape arguments.  By assuming smooth, analytic behavior in the
manifolds' thermodynamic functions, Shapir found that the
susceptibility (for a surface of dimension $D$) is proportional to the
displacement of the manifold in the limit of small field, which is
applied uniformly to the whole manifold, and $\chi_{D} \sim d^2 E /
dh^2 \sim L^{\theta +2 \tilde{\alpha}}$. $\tilde{\alpha}$ results from
an argument concerning the energy gap and the external field: $\Delta
E \sim h L^\zeta L^D$, where the external field couples to a droplet
of size $L^\zeta$ and the area of the manifold is $L^D$. This should
be equal to the energy difference: $\Delta E \sim
L^{\theta+\tilde{\alpha}}$, and thus $\tilde{\alpha} = 2 - \zeta$,
since $\theta =2\zeta+D-2$.  Hence the susceptibility was derived to
be $\chi_D \sim L^{D+2}$ and the susceptibility per unit hyper-surface
vary as $L^2$ for a manifold of scale $L$. This surprisingly is {\it
independent} of the type of the pinning randomness.

The landscapes of DP's have also been studied by J\"ogi and
Sornette~\cite{Joe98} by moving both end points step by step (with
periodic boundary conditions) from $(0,z)$ and $(L,z)$ to $(0,z+1)$
and $(L,z+1)$, and finding the ground state at each step.  They
studied the sizes of the {\it avalanches} as in self-organized
systems, i.e., the areas the polymer covers when moving to the next
position. They found a power-law for the avalanche sizes $P(S)\, dS
\sim S^{-(1+m)} \, dS$, where $m =2/5$, up to the maximum size
$S_{max} \sim L^{5/3}$. $S_{max}$ is actually the size of the largest
fluctuation of the polymer in such a controlled movement, i.e., $L
\times L^{\zeta}$. The authors did not however study the energy
variations of the manifold during the ``dragging process''.

In the limit of a large $h$ the Hamiltonian of Eq.~(\ref{H}) is
applicable to (complete) wetting interface in random systems with
three phases. In this case the manifold is the interface between i)
non-wet and ii) wet phases near iii) a hard wall at $z=0$.  This
problem was studied by mean-field arguments by first Lipowsky and M.\
Fisher and later elaborated numerically~\cite{Lip86,Wuttke91}. The
choice of wetting potential that corresponds to the linear field used
here is called in the wetting literature the weak-fluctuation regime
(WFL). In the WFL regime one can derive the wetting exponent $\psi$ by
a Flory argument, since the mean distance of the interface from the
inert wall $\overline{z}$ is of the order of the interface transverse
fluctuations $\xi_\perp$, see Fig.~\ref{fig1}, i.e., the field is
strong enough to bind the interface to the wall.  The Hamiltonian can
now be minimized by estimating the total potential to be $V_{tot} =
V_{fl}(\overline{z})+ V_W(\overline{z})$. $V_W \sim h z$ is the
wetting potential induced by the external field $h$.  The
fluctuation-induced potential $V_{fl}$ is the potential of the free
interface $V_{el} + V_r = \frac{\Gamma}{2} \{ \nabla z({\bf x}) \}^2 +
V_r ({\bf x},z)$, which gives, when minimized, $\xi_\perp \simeq
\xi_\|^\zeta$.  The elastic energy term $V_{el}$ gives the scaling of
$V_{fl}$ with respect to the correlation lengths, and expanding the
square-term of the gradient of the height gives $\rm{const}_1 +
\rm{const}_2 \left| \frac{\xi_\perp}{\xi_\|} \right |^2 +$ higher
order terms. So, $V_{fl} \sim \xi_\perp^{-\tau}$, where the decay
exponent $\tau = (2-2\zeta)/\zeta$ defines the scaling between the
pinning and elastic energies. Minimizing $V_{tot}$ in WFL, where
$\overline{z} \simeq \xi_\perp$ and is much larger than a lattice
constant, gives
\begin{equation}
\overline{z} \sim h^{-1/(1 +\tau)} \sim h^{-\psi} . 
\label{cwexp}
\end{equation} 
Thus the wetting-exponent $\psi = 1/(1+\tau)$ becomes
\begin{equation}
\psi = \frac{\zeta}{2-\zeta}.
\label{psi}
\end{equation} 
Huang {\it et al.}~\cite{Hua89,Wuttke91} did numerical calculations
for directed polymers at zero temperature using transfer-matrix
techniques for slab geometries, $L \gg L_z$, where $L$ is the length
of the polymer, and $L_z$ is the height of the system, confirming
roughly the expected exponent $\psi \simeq 0.5$. Our results,
presented in Section~\ref{wetting} are in line with the expectations
in (1+1) and we also present the first studies in (2+1) dimensions.

\section{Glassy energy landscapes: length scale of perturbations}
\label{glassy}

We next employ a version of the scaling arguments by Hwa and Fisher,
and M\'ezard for the next optimal position of the directed polymer.
The goal is to investigate the preferred length scale of the change
that takes place, i.e., the size of the ``droplet'' created with the
field. The numerical calculations presented in this paper are
done such a way, that first the ground state is searched and the
configuration stored. After that the external field is applied and
the energy is minimized again.  If there is a change in the
configuration, the difference in the mean height of the manifold, the
so called ``jump size'' is analyzed. Most of the studies are done in
isotropic systems, i.e., the height $L_z$ of the system is of the
order of the linear length of the manifold, $L_z \propto L$.
One finds that the behavior of the polymer has a {\it first
order character}: the optimal length scale is such that the whole
configuration changes, and this holds for higher dimensional
manifolds, too. 

Hwa and Fisher defined the next optimal position of the DP fixed at
one end and with displacement $\Delta$ to scale as $\Delta \sim
L^{\zeta}$, where $L$ is the length of the polymer. Now one assumes
that the energy {\it gaps} between two copies of the polymers with
overlap, and energies $E_0$ and $E_1$, grows as $E_1 -E_0 \sim
L^\theta$, i.e., is just energy fluctuation exponent of a polymer (as
demonstrated by M\'ezard numerically). Then as a generalization with
$n=1$ it follows that $E_1 -E_0 \sim \Delta^{\theta/\zeta}$.  The
external field has a contribution for the energy differences of
interfaces with the dimension $D$ $E_1 -E_0 \sim h L^D \Delta \sim h
\Delta^{1+D/\zeta}$.  Assuming that this difference balances the gap,
and using the hyperscaling law for the roughness and energy
fluctuation exponents $\theta = 2\zeta+D-2$, we get
\begin{equation}
h \sim \Delta^{\overline{\alpha}} = \Delta^{(\zeta-2)/\zeta}.
\label{excitation}
\end{equation}
The exponent $\overline{\alpha}$ is negative assuming that the
roughness exponent is below two, which is of course satisfied in the
case of both RF and RB domain walls.  Hence, the smaller the field the
larger the excitations and thus large excitations are the preferred
ones, at least below the upper critical dimension.  Besides RB systems
Eq.~(\ref{excitation}) works only for RF systems with a ferromagnetic
bulk (among others, the relation $E_1 -E_0 \sim L^\theta$ has not been
demonstrated to hold for RF interfaces). If the bulk of a RF magnet is
paramagnetic, the stiffness of DW's is expected vanish on large enough
length scales, translational invariance is broken, and $\Delta E =
E_1-E_0$ does not scale.  The scaling relation (\ref{excitation})
should hold also for general $n$. One should note in the case $h$ is
applied in one direction but the droplet may extend in $n$ dimensions,
in order to Eq.~(\ref{excitation}) to hold, also the {\em projection}
of the droplet in the applied field direction has to have the scaling
as $\Delta^{\theta/\zeta}$.

\begin{figure}[f]
\centerline{\includegraphics[width=9cm]{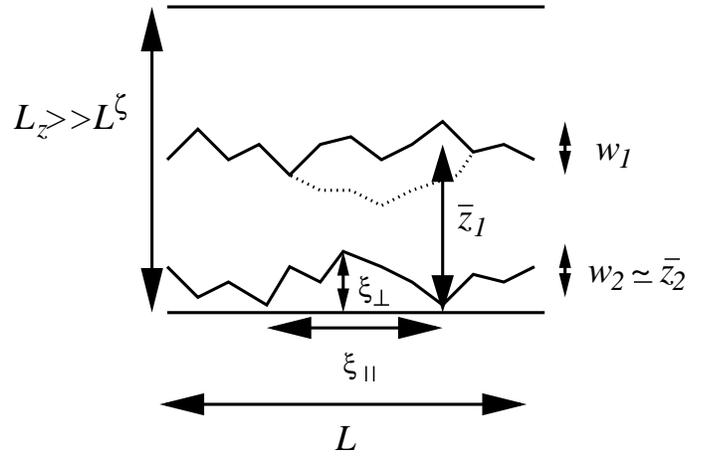}}
\caption{A schematic figure of two interfaces in a system of parallel 
length $L$ and perpendicular one $L_z > L^\zeta$ at heights
$\bar{z}_1$ and $\bar{z}_2 \simeq w_2$, with the corresponding
roughness values $w_1$ and $w_2$.  $\xi_\perp$ denotes the transverse
correlation length of a part of the interface of length $\xi_\|$. See
the text.  A droplet is seen in the interface of height $\bar{z}_1$ as
a dotted line.}
\label{fig1}
\end{figure}

\begin{figure}[f]
\centerline{\includegraphics[width=7cm,angle=-90]{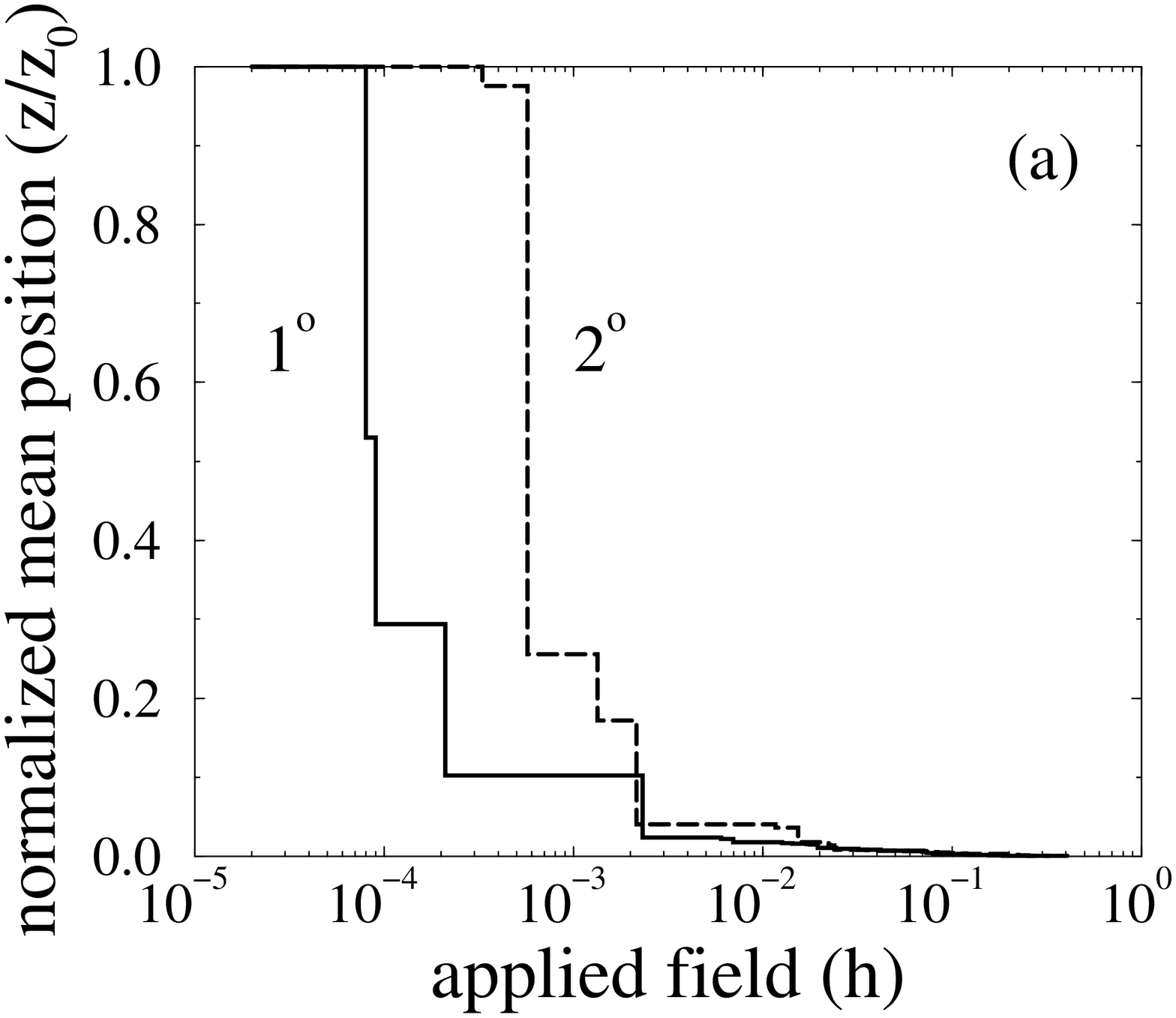}}
\centerline{\includegraphics[width=7cm,angle=-90]{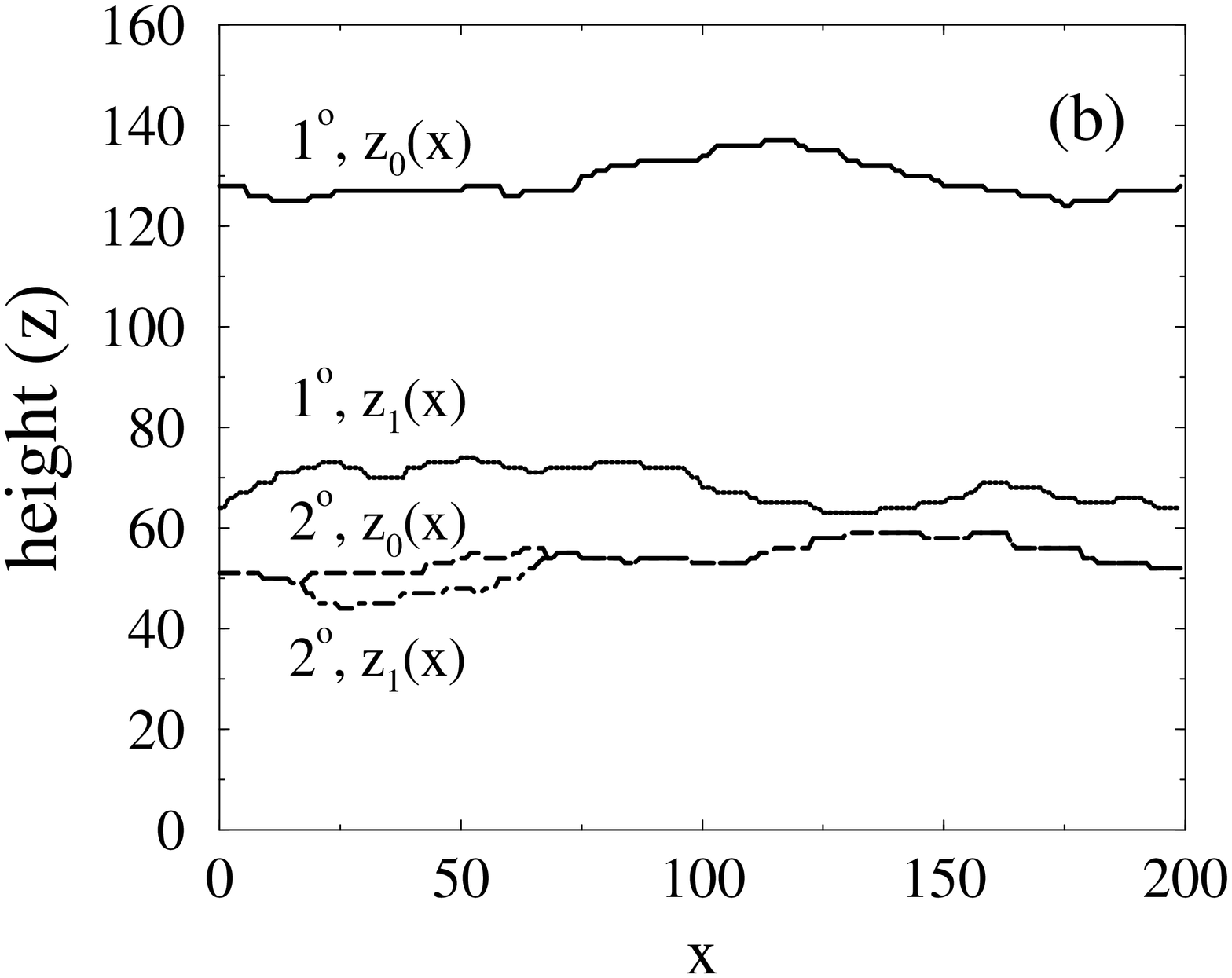}}
\caption{(a) Examples of two realizations of changes in mean heights
$\bar{z}$ of interfaces normalized by their original (global minimum)
positions $\bar{z}_0$ vs. applied field $h$ for $(1+1)$ dimensional
systems. The change in the external field is done in steps of $\Delta
h = 10^{-5}$.  Note the large jumps in both cases. $L^2=200^2$. $J_{ij,z}
\in [0\--1]$ uniform distribution and $J_{ij,x} = 0.5$ (random bond
disorder). (b) The expected scenarios (droplet formation, jump to the
lower edge of the system) before and after the first moves from global
minima $z_0(x)$ to $z_1(x)$.}
\label{fig2}
\end{figure}

In Fig.~\ref{fig2} it is shown what happens for two different random
realizations of DPs, when a perturbing external field is
applied. Fig.~\ref{fig2}(a) shows the mean height of the polymer
normalized by its original height $\overline{z}/\overline{z}_0$, when
the external field is increased. The first realization $1^\circ$ shows
a large {\it jump} of a size half of the height of its original
position at {\it first jump field} $h_1 = 8 \times 10^{-5}$.  In
Fig.~\ref{fig2}(b) the two positions of $1^\circ$, before and after
the first jump, $z_0(x)$ and $z_1(x)$, are shown.  The other scenario
would be that the directed polymer would continuously undergo small
geometric adjustments and get meanwhile bound to the wall, $z=0$.  An
example of such a droplet, is shown in Fig.~\ref{fig2}(b) as the
second realization $2^\circ$, see also Fig.~\ref{fig1}.  However, a
further field increase, demonstrated in Fig.~\ref{fig2}(a) for
$2^\circ$, leads to a global jump too.  This leads to a picture in
which assuming a starting position far enough from the system
boundary, a finite number of large jumps exists from the original
position $z_0({\bf x})$ to the positions $z_1({\bf x}), z_2({\bf x}),
\ldots, z_n({\bf x}), \ldots $, closer and closer to the wall, i.e.,
$z_0({\bf x}) > z_1({\bf x}) > z_2({\bf x}) > \ldots > z_n({\bf x}) >
0$.

The mean jump length is defined to be $\Delta z_n = \overline{z}_{n-1}
- \overline{z}_{n}$, and the jumps take place at the fields $h_1, h_2,
\ldots, h_n, \ldots$. The field value $h_1$ corresponds the jump from 
$z_0({\bf x})$ to $z_1({\bf x})$. Finally, after a finite number of
jumps, with the interface being in the proximity of the wall, around $h
\simeq 2 \times 10^{-2}$ in Fig.~\ref{fig2}(a), see also
Fig.~\ref{fig1}, the interface evolves inside the last valley next
to the wall continuously, and the wetting behavior, Eqs.~(\ref{cwexp})
and (\ref{psi}), applies. We discuss this picture of consequent jumps
before the wetting regime in Section~\ref{first_jump} together with
the numerical results.

\begin{figure}[f]
\centerline{\includegraphics[width=7cm]{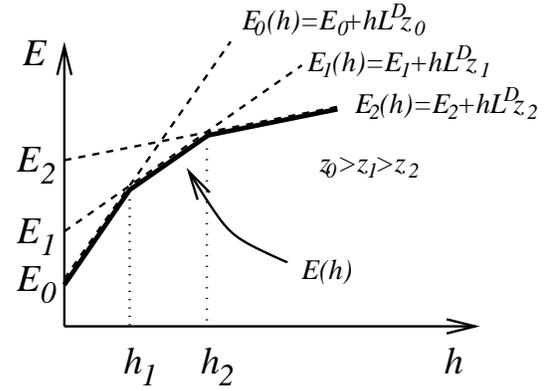}}
\caption{The level-crossing phenomenon 
for interfaces in random systems in the presence
of an external field.  Originally the interface lies at height $z_0$ 
and has an energy $E_0$. When the field is applied its energy increases
linearly 
and at $h_1$ the interfaces jumps to $z_1 < z_0$, with the
energy $E_1(h_1)=E_1
+ h_1 L^D z_1$, where $E_1$ is the energy of the interface at $z_1$
without the field. Similar behavior takes place at $h_n$, $n=2,3,\dots $,
when the interface moves from $z_{n-1}$ to $z_n < z_{n-1}$. The thick line
represents the 
minimum energy of the interface $E(h)$ as a function of the
field $h$, with discontinuities of the derivative at $h_n$.}
\label{fig3}
\end{figure}

The global changes (large jumps) induce finite changes in the
magnetization ($m$), and are reminiscent of first order phase
transitions. Note that the field $h_1$ of the first change or jump
obeys for any particular ensemble a probability distribution, and thus
a co-existence follows between systems that have undergone a jump or
change in $m$ and those that are still in the original state. This
``transition'' is a result of level-crossing between the valleys in
the energy landscape for the interface, sketched in
Fig.~\ref{fig3}. To change between the geometrically separated minima,
an external field is applied, which at jumps plays the role of a
latent heat.  Originally the interface lies at height $\overline{z}_0$
and has an energy $E_0$. When the field is applied interface's energy
increases by $h L^D \overline{z}_0$ and at $h_1$ the interface jumps
to $\overline{z}_1 < \overline{z}_0$ having energy $E(h_1)=E_1 + h_1
L^D \overline{z}_1 = E_0 + h_1 L^D \overline{z}_0$, where $E_1$ is the
energy of the interface at $z_1$ without the field. Similar behavior
takes place at $h_n$, $n=2,3,\dots $, when the interface moves from
$\overline{z}_{n-1}$ to $\overline{z}_n < \overline{z}_{n-1}$, until
the cross-over to the wetting regime.

\begin{figure}[f]
\centerline{\includegraphics[width=7cm,angle=-90]{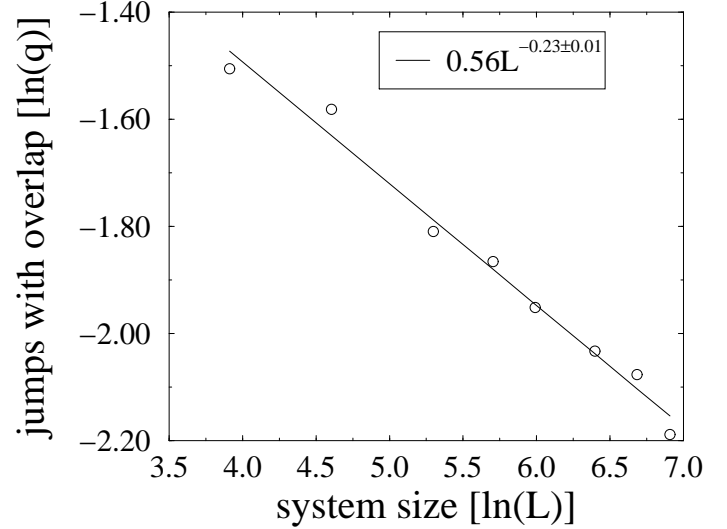}}
\caption{The fraction of jumps of interfaces with initial
global minimum in an arbitrary position that have an overlap between
interfaces before and after jump $q$ as function of the system sizes
$L$. The system sizes are $L^2 = 50^2\--1000^2$
and the number of realizations $N=3000\--5000$. The line is a
 least-squares
fit to the data with the power-law behavior $L^{-0.23 \pm 0.01}$.}
\label{fig4}
\end{figure}

In the thermodynamic (TD) limit, i.e., for very large systems, when
$L, L_z \to \infty$, the interface may originally be located anywhere
in the system. Assuming that the roughness exponent $\zeta$ does not
define only the width of the manifold, but also the width of the
minimum energy valley, i.e., $L^\zeta$ is the only relevant length
scale in the transverse direction, gives that the manifold should have
$N_z \sim L_z/L^\zeta$ minima from which to choose the global minimum
position. For $\zeta <1$ the number of the minima grows with system
size, if the geometry is kept isomorphic, $L_z \propto L$.  The
requirement $\zeta <1$ holds for all RB interfaces; and for RF ones
with ferromagnetic bulk, when $D>1$. In the special case that the
local minimum closest to the wall coincides with the global minimum,
the wetting regime is entered at once without any jumps. The
prediction of Eq.~(\ref{excitation}) that large scale excitations
should dominate is an asymptotic one, and for finite system sizes one
may find also small excitations, which are less costly in energy
(Fig.~\ref{fig2}).  However, the fraction of these cases decreases
with system size, see Fig.~\ref{fig4}. It depicts in $d=2$ up to $L^2
= 1000^2$, with at least $N=3000$ realizations per data point, the
probability of finding a non-zero overlap $q$ between the interfaces
before and after the first jump. $q$ is the disorder average of the
fraction of the samples, which have overlap between the cases before
and after the jump, i.e., for at least one $x$ $z_0(x) = z_1(x)$.  The
probability goes as $q \simeq 0.56L^{-0.23 \pm 0.01}$, and confirms
the expected behavior that for isotropic systems in the $L \to \infty$
limit the first jump is always without an overlap, i.e., a macroscopic
one.  A first guess would give that $q \sim 1/N_z \sim L^{\zeta-1}$.
A similar trend seems to exist in $(2+1)$ dimensions, too, but our
statistics is not good enough to determine the exponent of $q$
reliably. This is since $q$ is averaged over a binary distribution
(overlapping and non-overlapping jumps), and thus very large ensembles
would be needed.

\section{Extreme statistics arguments}
\label{argumentti}

Next we compute the field $h_1$ and the susceptibility,
Eq.~(\ref{Eqsuskis}), by taking into accont the ``co-existence''
of the original ground state and the state after the jump.  Assuming
that the relevant process in the response of a domain wall is the
large-scale jump, the susceptibility per spin of a system with a
domain wall, Eq.~(\ref{Eqsuskis}), may be written in the form
\begin{equation}
\chi = \lim_{h \to 0+} \left \langle \frac{\Delta m(h)}{\Delta h} 
\right \rangle \simeq \left \langle \frac{\Delta z_1}{L_z} 
\right \rangle \lim_{h \to 0+} P(h_1),
\label{Eqsuskis2}
\end{equation} 
where $P(h_1)$ is the probability distribution of the first jump
fields with the corresponding first jump size $\Delta z_1$, and the
magnetization $m(h) \sim z(h)/L_z$.  The limit $h \to 0+$ is taken
over the probability of having a jump, and thus the susceptibility
will reflect the co-existence phenomenon discussed above. One uses
also a plausible assumption, found to be true in the numerics, that
the change in the interface does not depend on the threshold field
$h_1$.  A further simplification is obtained by assuming (this will be
shown numerically later) that the jumps have an invariant size
distribution, independent of $L$ and $L_z$, after a normalization with
$L_z$.  This gives $\langle \Delta z_1 \rangle \sim L_z$. Hence the
finite size scaling of the susceptibility per spin depends only on the
probability distribution of the first jump field in the limit of the
vanishing external field, and in particular on the ``rare events''
measured by such a distribution. The dependence of $h_1$ on the
anomalous scaling is found to be true not only for directed polymers
but also for the higher dimensional case.  Such behavior is contrary
to Shapir's work that assumed smooth, analytic thermodynamic functions
and it is thus no surprise that the scaling of the susceptibility,
derived in detail in Section~\ref{results}, differs from that.

Next we first derive the distribution for the first jump field and its
relation to the lowest energy gap probability distributions.  Then the
probability distribution and the finite size scaling of the lowest
gaps and also the lowest energy level are derived using extreme
statistics arguments (see also \cite{Galambos,DBL} and our shorter
account of parts of this work \cite{unpublished}). These gaps have a
logarithmic dependence on the number of local minima.

\begin{figure}[f]
\centerline{\includegraphics[width=8cm]{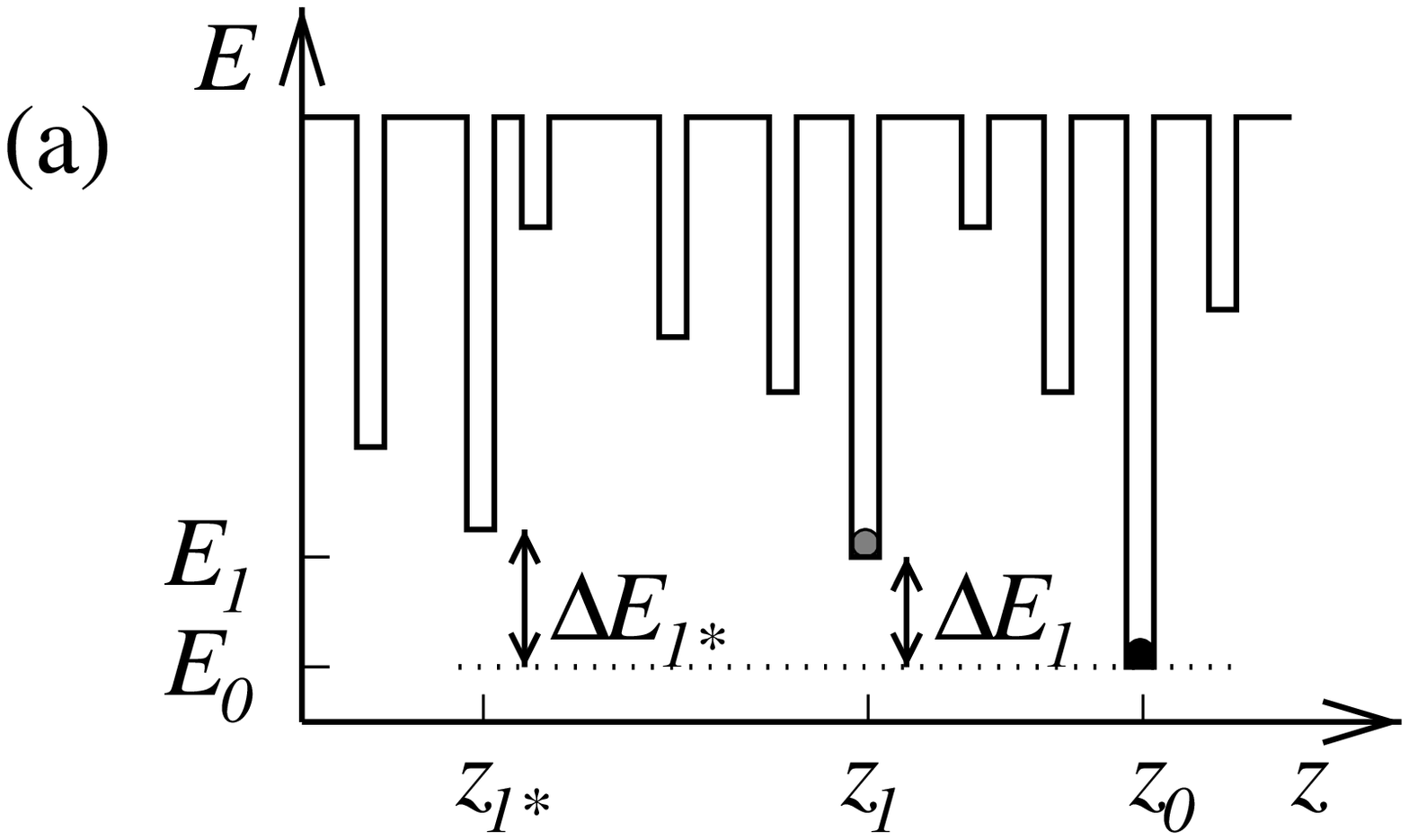}}
\centerline{\includegraphics[width=8cm]{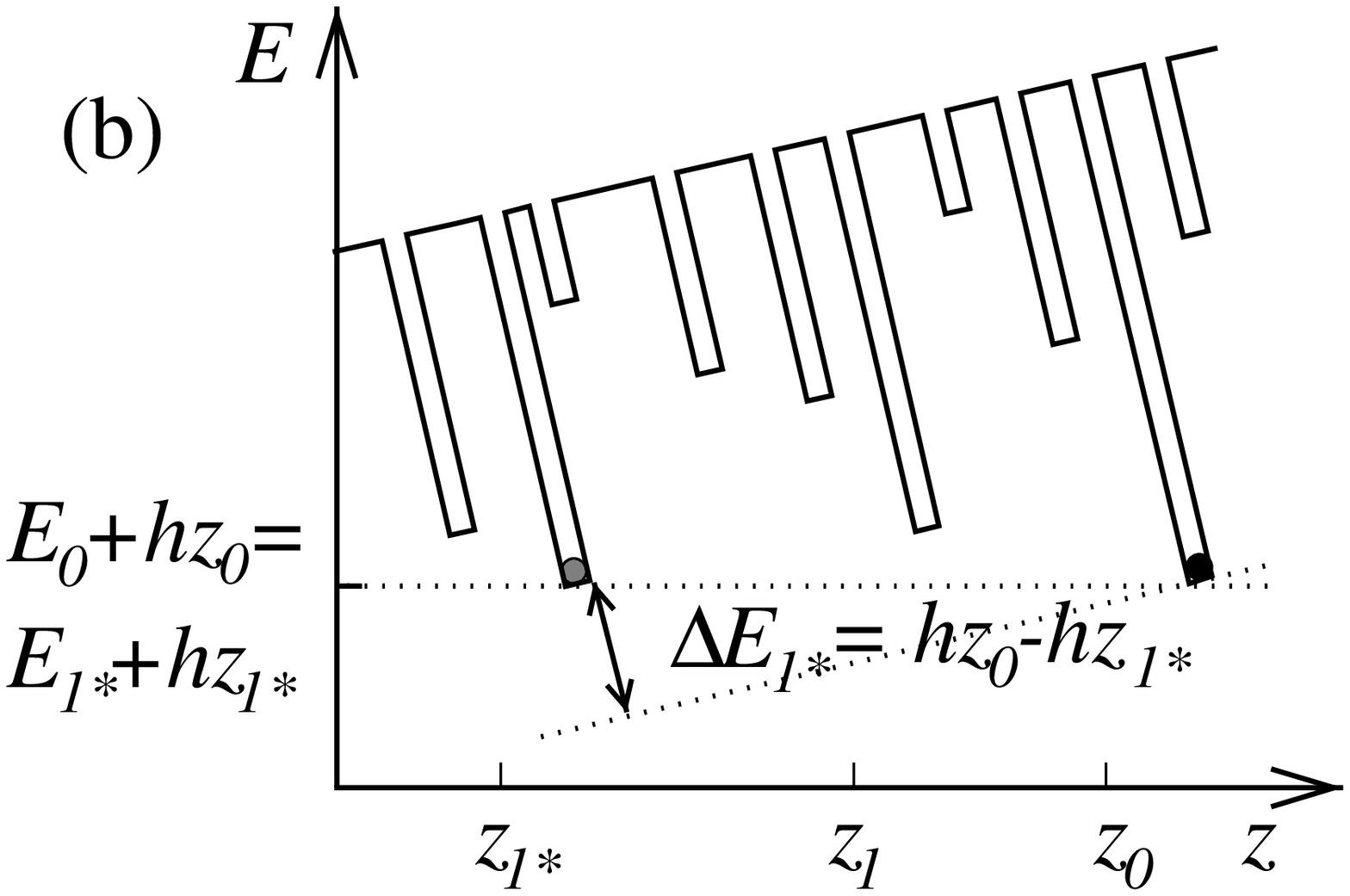}}
\caption{(a) A simplified view of the minima in the energy 
landscape of a random system. $\Delta E_1 = E_1 -E_0$ is the energy
gap between the ground state at $z_0$, denoted with a black
circle, and the second lowest minimum at $z_1$ denoted with a gray
circle. $\Delta E_{1^*} = E_{1^*} -E_0$ is the energy gap
(normalized with $L^D$) between the ground state and the minimum at
$z_{1^*}$. (b) The view of the minima in the random system when the
field is applied. At $h_1$ the interface moves from $z_0$ to
$z_{1^*}$, indicated with a gray circle, since the energy difference
$\Delta E_{1^*} = E_{1^*} -E_0 = h_1 z_0 - h_1 z_{1^*} = 0$ while
all the other $\Delta E$'s are greater. However, often $E_{1^*}$ and
$E_1$ coincide.}
\label{fig5}
\end{figure}

In order to derive the probability distribution of the first jump field
$P(h_1)$, let us first calculate the probability $P_0$, that with a 
certain test field $h'<h_1$, the interface has not changed yet, i.e.
\begin{equation}
P_0(h_1 >h')= 1-P_0(h_1 \leq h') = 1-\int_0^{h'} P(h) \, dh.
\label{P0}
\end{equation} 
By differentiating $P_0(h_1 >h')$, one gets the probability distribution 
of the first jump field, since
\begin{equation}
P(h_1) = \frac{\partial}{\partial h} \left.  P_0(h_1 \leq h') 
\right|_{h'=h_1}.
\label{P0derv}
\end{equation} 
We assume all the minima to be non-correlated and well separated from
each other, see Fig.~\ref{fig5}(a).  There we have a global energy
minimum $E_0$ at $z_0$ (note, that here the energy values as well as
the field contributions are normalized by $L^D$, constant with a fixed
system size), and energy gaps $\Delta E_1 = E_1 - E_0$ and $\Delta
E_{1^*} = E_{1^*} - E_0> \Delta E_1$ with energies $E_1$ at $z_1$ and
$E_{1^*}$ at $z_{1^*}$, respectively. Then we apply the field $h$,
i.e., we tilt the energy landscape, see Fig.~\ref{fig5}(b). Due to the
statistical tilt symmetry the external field $h$ is assumed not to
change the shape of the random landscape.  At the lowest tilt to move
the interface, i.e., at field $h_1$, it moves to $z_{1^*}$, since with
the field $h_1$ $E_0(h_1) = E_0 +h_1z_0 = E_{1^*}(h_1)
=E_{1^*}+h_1z_{1^*} < E_1(h_1) = E_1+h_1z_1$.  However, sometimes $E_1$
and $E_{1^*}$ coincide, so that the interface jumps to the true second
lowest minimum.  Now we index the positions of the $N_z$ minima as
$z_i$ (need not to be in order), so that $h(z_0-z_i) = h \Delta z_i$,
and we have
\begin{eqnarray}
P_0(h_1>h') = \prod_{i=1}^{N_z} {\rm prob}_i(\Delta E_i > h\Delta z_i) 
\nonumber \\
= \prod_{i=1}^{N_z} \left\{ 1- \int_0^{\Delta z_ih} \hat{P}_i(\Delta E_i) \, 
d(\Delta E_i)  \right\}.
\label{P0prod}
\end{eqnarray}
By assuming all the gap energies from the same probability
distribution $\hat{P} (\Delta E)$, setting $\Delta z_k \hat{=}
\frac{k}{N_z}$ and taking the continuum-limit we get for the
probability distribution of the first jump field using
Eqs. (\ref{P0}), (\ref{P0derv}) and (\ref{P0prod}),
\begin{eqnarray}
P(h_1) = \exp\left[-\int_1^{N_z} \int_0^{kh_1/N_z}
\hat{P}(\Delta E) \, d(\Delta E) \, dk \right] \times \nonumber \\
\int_1^{N_z} \frac{ \hat{P}(kh_1/N_z)}{1 -\int_0^{kh_1/N_z} 
\hat{P}(\Delta E) \, d(\Delta E)} \, dk.
\label{Ph1}
\end{eqnarray}
There are two approaches to go on from the distribution of
Eq.~(\ref{Ph1}): one may either compare its prediction with numerics
with a trial distribution for $\hat{P}$, or alternatively proceed by a
complete Ansatz for the valley energies, from which the gap naturally
follows. We now attempt the latter, and comment on the first one
together with the numerics.

In order to calculate the probability distribution for the gap
energies $\hat{P}(\Delta E)$, we use a probability distribution for
the energy minima suitable for directed polymers, which has the
advantage of decaying faster than any power-law, ${\mathcal P}(E) \sim
\exp(-BE^\eta)$, $\eta>0$.  For directed polymers with the ``single
valley'' boundary condition case, i.e., one end of the manifold fixed
and thus no room for other minima, it is known mostly numerically that
the bulk of the distribution of the energy of a directed polymer
${\mathcal P}(E)$ is Gaussian (the exponent $\eta =2$ in the
exponential), except for the tails, ($E <E_{min}$ and $E > E_{max}$)
in which different cut-offs, ($\eta_- =1.6$ and $\eta_+=2.4$) take
over~\cite{KBM,HaH95}. For the other manifolds that one might be
interested in the actual distributions have not generally been
studied.  In the following we usually approximate the distribution
with Gaussian distribution (this is valid for $1/N_z \geq E_{min}$ and
$< E_{max}$).  The validity of the approximation is discussed in
Section~\ref{results}, when the analytical arguments are compared with
the numerics.

The distribution ($\eta =2$ for Gaussian) for ${\mathcal P} (E)$ is
written in the form:
\begin{equation}
{\mathcal P}(E) =  k \exp\left \{-\left( {|E - \langle E 
\rangle |\over \Delta E} \right)^\eta  \right \},
\label{distr}
\end{equation}
where $\langle E \rangle \sim L^D$ is the average energy of the
manifold and $k \sim (\Delta E)^{-1} \sim L^{-\theta}$
normalizes the distribution.  The extreme statistics argument
goes so that in a system with 
\onecolumn 
$N_z \sim L_z/L^\zeta$ minima the
probability for the lowest energy to be $E$ is
\begin{equation}
L_{N_z}(E) = N_z {\mathcal P}(E) \left \{1-C_1(E) \right \}^{N_z-1},
\label{LNdistr}
\end{equation}
where $C_1(E) = \int_{-\infty}^E {\mathcal P}(e) \, de$ is called the
error-function when $\eta =2$.  For Gaussian ${\mathcal P}(E)$,
likewise for general $\eta>0$, $L_{N_z}(E)$ is known to be Gumbel
distributed, $\exp(u-\exp u)$~\cite{boumez,Gumbel,Rowlands}. The gap
$\Delta E_1$ follows similarly as $L_{N_z}(E)$.  Its distribution,
$G_{N_z}(\Delta E_1,E)$ is given by
\begin{eqnarray}
G_{N_z}(\Delta E_1,E) = {N_z(N_z-1)\over 2} {\mathcal P}(E) 
{\mathcal P}(E+\Delta E_1) \nonumber\\
\{1-C_1(E+\Delta E_1)\}^{N_z-2}.
\label{DeE}
\end{eqnarray}
$G_{N_z}(\Delta E_1,E)$ is the probability that if the lowest energy
manifold has an energy $E$, then the gap to the next lowest 
energy level is $\Delta E_1$.  $G_{N_z}(\Delta E_1,E)$ is actually a
generalization of Gumbel distribution. Integrating Eq.~(\ref{DeE})
over all energies gives the probability distribution for the $\Delta
E_1$
\begin{eqnarray}
\hat{P}(\Delta E_1) = \int_{-\infty}^{\infty} G_{N_z}(\Delta E_1,E) \, dE.
\label{DE1}
\end{eqnarray}
This becomes for Gaussian ($\eta=2$) and $\Delta E_1 \ll \langle E \rangle$
\begin{eqnarray}
\hat{P}(\Delta E_1) = \int_{-\infty}^{\infty} {N_z(N_z-1)\over 2} k^2 
\exp \left \{ {- (E- \langle E \rangle)^2 -2\Delta E_1(E- \langle E 
\rangle) \over \Delta E^2 } \right \} 
\left \{ k\, {\rm erfc} \left( E +\Delta E_1- \langle E \rangle \over 
\Delta E \right) \right\}^{N_z-2} \, dE,
\label{DE1_2}
\end{eqnarray}
where erf denotes the error-function and erfc $=1-$ erf.
Neglecting all ${\mathcal O} \left[ \left\{ k\, {\rm erfc} \left( E
+\Delta E_1 - \langle E \rangle \over \Delta E \right) \right\}^2
\right]$ and higher order terms and using a Taylor expansion around $\Delta
E_1=0$, i.e., Maclaurin-series, gives
\begin{eqnarray}
\hat{P}(\Delta E_1) = \int_{-\infty}^{\infty} {N_z(N_z-1)\over 2} k^2 
\exp \left \{ {- (E- \langle E \rangle)^2 -2\Delta E_1(E- \langle E 
\rangle) \over \Delta E^2 } \right \} \nonumber \\
\left[ 1 -(N_z-2) k\, {\rm erf} \left( E - \langle E \rangle \over \Delta E 
\right) +(N_z-2) \Delta E_1 k \exp \left \{ {- (E- \langle E \rangle)^2 \over 
\Delta E^2 } \right \} \right ] \, dE.
\label{DE1_3}
\end{eqnarray}
\twocolumn  
The first two terms within the second parenthesis
dominate for $\Delta E_1$ small, and thus we see that to
first order of $\hat{P}(\Delta E_1) \sim$ const (compare
with the numerics presented below). In particular,
one should notice that the
probability distribution does not vanish for $\Delta E_1 \simeq 0$.

This approach is very similar to the extreme statistics calculation of
the one-dimensional version of the random energy model
by Bouchaud and M\'ezard~\cite{boumez} (see
Section~\ref{thermo}). They derived the probability distribution for
the location that gives minimum energy for the system, with the
difference to our case that Hamiltonian reads $\frac{\gamma}{2}x^2 +
E(x)$.  The result becomes such that the
distribution for the position is approximately Gaussian, too.  The
one-dimensional system is close to our example except for the
functional form of the external potential which is quadratic instead
of a linear one. There is however one essential difference in that in
their analysis the ``field value'' is fixed, whereas in our case we
are interested in what happens in any particular sample as the field
is varied. Nevertheless such a calculation could be compared to the
distribution of the interface locations at a fixed field $h$.

More important than the actual distributions is that the finite size
scaling of the gap energies may be computed.  Let us start by
calculating the typical lowest energy level. The average of it is
given by
\begin{equation}
\langle E_0 \rangle = \int_{-\infty}^{\infty} E L_{N_z}(E) \,dE,
\label{E0true}
\end{equation}
and the typical value of the lowest energy is estimated from
\begin{equation}
N_z \frac{1}{k} {\mathcal P}(\langle E_0 \rangle) \approx 1.
\label{E0estimate}
\end{equation}
Note, that in approximating the integral, Eq.~(\ref{E0true}), with the
aid of the distribution, Eq.~(\ref{E0estimate}) (in the limit $C_1(E)$
in Eq.~(\ref{LNdistr}) is small) the normalization $1/k$ should be
taken into account.  Eq.~(\ref{E0estimate}) gives
\begin{equation}
\langle E_0 \rangle \sim \langle E \rangle - \Delta E \left 
[ \ln (N_z)  \right ]^{1/\eta},
\label{typicalene}
\end{equation}
where $\Delta E \sim L^\theta$ and for Gaussian $\eta =2$. 

To estimate the typical value of the gap, we make a similar
approximation as in Eq.~(\ref{E0estimate}) for $\langle E_0 \rangle$,
\begin{equation}
{N_z(N_z-1)\over 2 k^2} {\mathcal P}(\langle E_0 \rangle ) 
{\mathcal P}(\langle E_0\rangle + \langle \Delta E_1 \rangle ) \approx 1,
\end{equation}
which with (\ref{typicalene}) and the fact that $| \langle \Delta E_1 
\rangle |  \ll | \langle E_0 \rangle| $ yields,
\begin{equation}
\langle \Delta E_1 \rangle \approx { \Delta E^\eta \over \eta
(\langle E \rangle - \langle E_0\rangle)^{\eta-1}} 
\approx { \Delta E \over \eta \left [ \ln(N_z) \right ]^{(\eta-1)/\eta}}.
\label{gap}
\end{equation}
We thus find that the gap scales as $\Delta E_1 \sim \Delta E [\ln
(N_z)]^{-1/2} \sim L^{\theta} [\ln (L_z L^{-\zeta})]^{-1/2}$, when
$\eta =2$ and assuming as in the previous section, that $N_z \sim
L_z/L^\zeta$. If the interfaces are flat, i.e., $\zeta =0$,
which is true for  $\{100\}$ RB interfaces below
the cross-over roughening scale $L_c$
\cite{Bouchaud92,Alava96,Raisanen98,unpublished2} if randomness
is weak, the same arguments should
hold. However, then the energy distribution is pure Gaussian, $\eta
=2$, and $\theta =D/2$ due to Poissonian statistics.

\section{Susceptibility of manifolds}\label{suskis}

\subsection{Numerical method}\label{numerics}

For the numerical calculations the RB Hamiltonian (\ref{RHamilton})
with $J_{ij} > 0$, $H=0$, has been transformed to a random flow
graph. The graph is formed by the Ising lattice and two extra sites:
the source and the sink; and the coupling constants $2J_{ij} \equiv
c_{ij}$ between the spins correspond to flow capacities $c_{ij}
\equiv c_{ji}$ from a site $S_i$ to its neighboring one
$S_j$~\cite{Alavaetal}.  The graph-theoretical optimization algorithm,
a maximum-flow minimum-cut algorithm, enables us to find the
bottleneck, which restricts the amount of the flow that can get from
the source to the sink given the capacities in such a random graph.
This bottleneck, a path $P$ which divides the system into two parts
(sites connected to the sink and sites connected to the source) is the
minimum cut of the graph and the sum of the capacities belonging to
the cut $\sum_P c_{ij}$ equals the maximum flow, the smallest of all
cut-paths in the system.  The source is connected to the sites in the
Ising lattice which are forced to be up and the sink is connected to
the sites which are forced to be down. The value of the maximum flow
is the total minimum energy of the domain wall, equivalent to the
minimum cut. The maximum flow algorithms can be proven to give the
exact minimum cut for all the random graphs, in which the capacities
are positive semidefinite and with a single source and
sink~\cite{network}.  The best known maximum flow method is by Ford
and Fulkerson and called the augmenting path method~\cite{FordF}. We
have used a more sophisticated method called push-and-relabel by
Goldberg and Tarjan~\cite{Goltar88}, which we have optimized for our
purposes.  It scales almost linearly, ${\mathcal O}(n^{1.2})$, with
the number of spins and gives the ground state DW in about minute for
a million of spins in a workstation. Notice that one could use for
DP's the usual transfer matrix method as well, but the max-flow
implementation is convenient in the case of systematic perturbations
to each sample.

In this study we have done simulations for $(1+1)$ and $(2+1)$
dimensional manifolds. The system sizes extend to $L \times L_z
=1000^2$ and $L^2 \times L_z = 400^2 \times 50$. The number of
realizations $N$ ranges from $200$ to $2 \times 10^4$. The random
bonds are such that in 2D $J_{ij,z} \in [0,1]$ uniform distribution
and $J_{ij,x} = 0.5$ unless otherwise mentioned. In three dimensions
we have either $J_{ij} \in [0,1]$ uniform distribution in all $x,y,$
and $z$ directions or dilution type disorder, $P(J_{ij}) = p \,
\delta(J_{ij}-1) + (1-p) \, \delta(J_{ij})$, i.e., a bond has a value
of unity or zero with the probability $p$. We have used $p=0.5$ and
$p=0.95$. The linear field contribution is applied in the
$z$-directional bonds as $J_\perp(z) = J_{random} + hz$. When the jump
field values with the corresponding jump distances are searched, the
precision is as small as $\Delta h =10^{-5}$ in order to be sure that
no smaller changes would occur between the jumps, which could be the
case if the precision was much larger. Periodic boundary conditions
are used in $x$ and $y$ directions.

When studying the susceptibility, we have controlled the number of the
minima in the systems to avoid fluctuations in $N_z$ (and the
grand-canonical ensemble).  To fix the number of the minima $\langle
N_z \rangle \sim L_z/L^\zeta$ in a certain system size, we have set
the initial position of the interface $\overline{z}_0$ in a fixed size
window at height $\overline{z}_0/L_z \simeq \rm{const}$. If the ground
state interface is originally outside the window, large enough only
for a single valley, it is discarded so that the statistics quoted are
based on the successful attempts. If the window is well-separated in
space from the $z$-directional boundaries it is obvious that this
sampling has no effect on the statistical properties as the average
energy.  After an original ground state is found, with an energy
$E_0$, the external field $h$ is applied by increasing the couplings
by constant steps until the first jump is observed~\cite{resgraph}.
We have also calculated the energy gap between the global minimum and
the next lowest minimum. In that case the initial position of the
interface $\overline{z}_0$ is also set to be in a fixed size window at
height $\overline{z}_0/L_z \simeq \rm{const}$ by discarding all the
samples with the global minimum outside the window. After that the lattice
is reduced so that bonds in and above the window are neglected and the
new ground state, its energy $E_1$, and the corresponding gap energy
$\Delta E_1$ are found. Although the discarding procedure is slow we
have at least $N=500$ realizations up to system sizes $L=300$,
$L_z=500$.
 
\subsection{Results}\label{results}

In order to compare the analytical arguments presented in
Section~\ref{argumentti} and in the end to compute the total
susceptibility of the manifolds we will first study numerically in
$(1+1)$ dimensions the finite size scaling of the average lowest
energy level. Then the shape of the energy gap distribution as well as
the finite size scaling of the average gap energies are considered.
After that the numerical results of the first jump field distribution
are presented together with the finite size scaling of the average
first jump field. The shape of the first jump field distribution and
the finite size scaling of its average are shortly reported for
$(2+1)$ dimensional interfaces, too. The susceptibility is derived
from the distribution and the finite size scaling of the first jump
field.  In the end of the section the jump distance distributions are
considered.

\begin{figure}[f]
\centerline{\includegraphics[width=7cm,angle=-90]{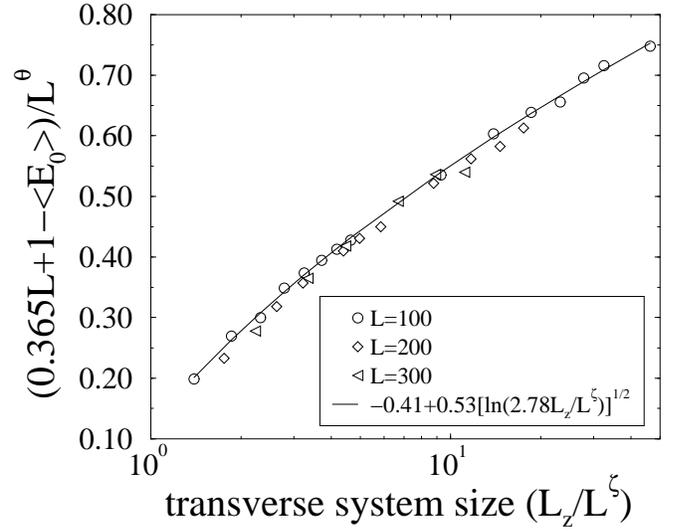}}
\caption{The scaling of the ground state energy, $(\langle E \rangle - 
\langle E_0 \rangle)/L^\theta$ vs. $L_z/L^{\zeta}$, see 
Eq.~(\ref{typicalene2}) and the text, for the system sizes
$L=$100, 200, and $300$, each with $\bar{z}_0/L_z \simeq
\rm{const}$. $\theta =1/3$, $\zeta =2/3$. The line $-0.41+0.53 [ \ln
(2.78 L_z/L^\zeta)]^{1/2}$ is a guide to the eye.  The number of
realizations ranges from $N=500$ for $L=300$, $L_z=500$ to $N=2000$
for $L=200$, $L_z=600$.}
\label{fig6}
\end{figure}

In order to see the logarithmic correction of the lowest energy level
$\langle E_0 \rangle$, Eq.~(\ref{typicalene}), when the height of a
system, and thus the number of the minima, are increased, we have
plotted in Fig.~\ref{fig6} for three different lengths of the directed
polymers $L=100,200,$ and $300$
\begin{equation}
\frac{\langle E \rangle - \langle E_0 \rangle}{L^\theta} \sim \left 
[ \ln  \left( N_z  \right) \right ]^{1/2}, 
\label{typicalene2}
\end{equation}
where $N_z \sim L_z/L^{\zeta}$, and $\langle E \rangle \simeq
0.365L+1$.  The prefactor, 0.365, of the average energy of a polymer
with only a single valley in a system, $\langle E \rangle$, we have
estimated by calculating systems of sizes $L_z \times L$, where $L_z =
6.5\times w$, $w$ is the average roughness of a polymer, up to system
sizes $50 \times 1000$ with $2000$ realizations. One should note, that
the prefactor is highly sensitive to the disorder type and boundary
conditions, c.f. Ref.~\cite{Hansenetal}, and the constant factor
(unity) also to the estimate of the size of the single valley. We have
used it as a free parameter. Although for the following results the
relevant part of the tail of $\langle E \rangle$ should not be a pure
Gaussian (at least if we have enough many minima $N_z$, so that the
bulk of the distribution is avoided), we have used for $1/\eta = 1/2$
instead of $1/\eta = 1/1.6=0.625$ in the fit of the logarithmic
correction. In practice one can not tell these two choices apart in
the range of the system sizes used.

\begin{figure}[f]
\centerline{\includegraphics[width=7cm,angle=-90]{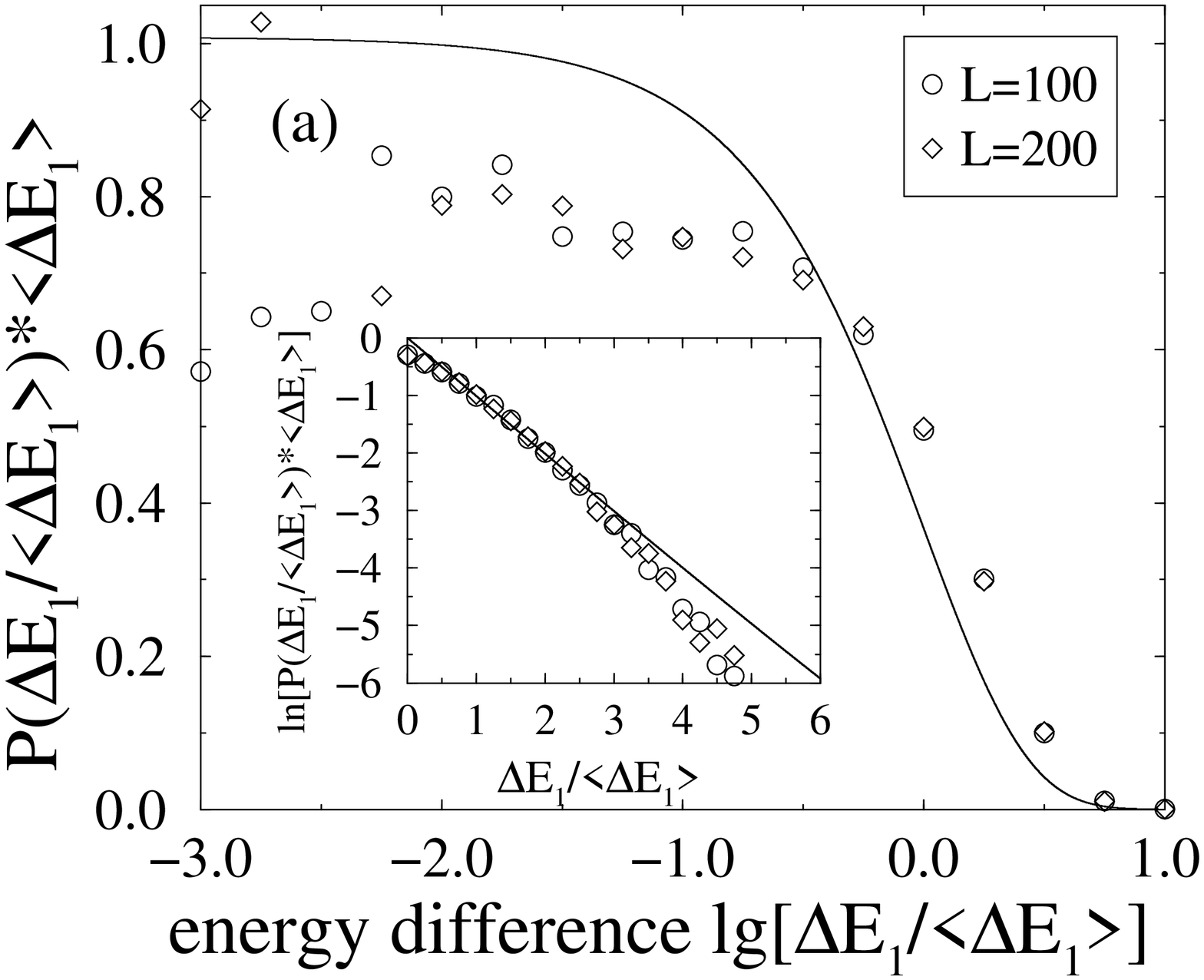}}
\centerline{\includegraphics[width=7cm,angle=-90]{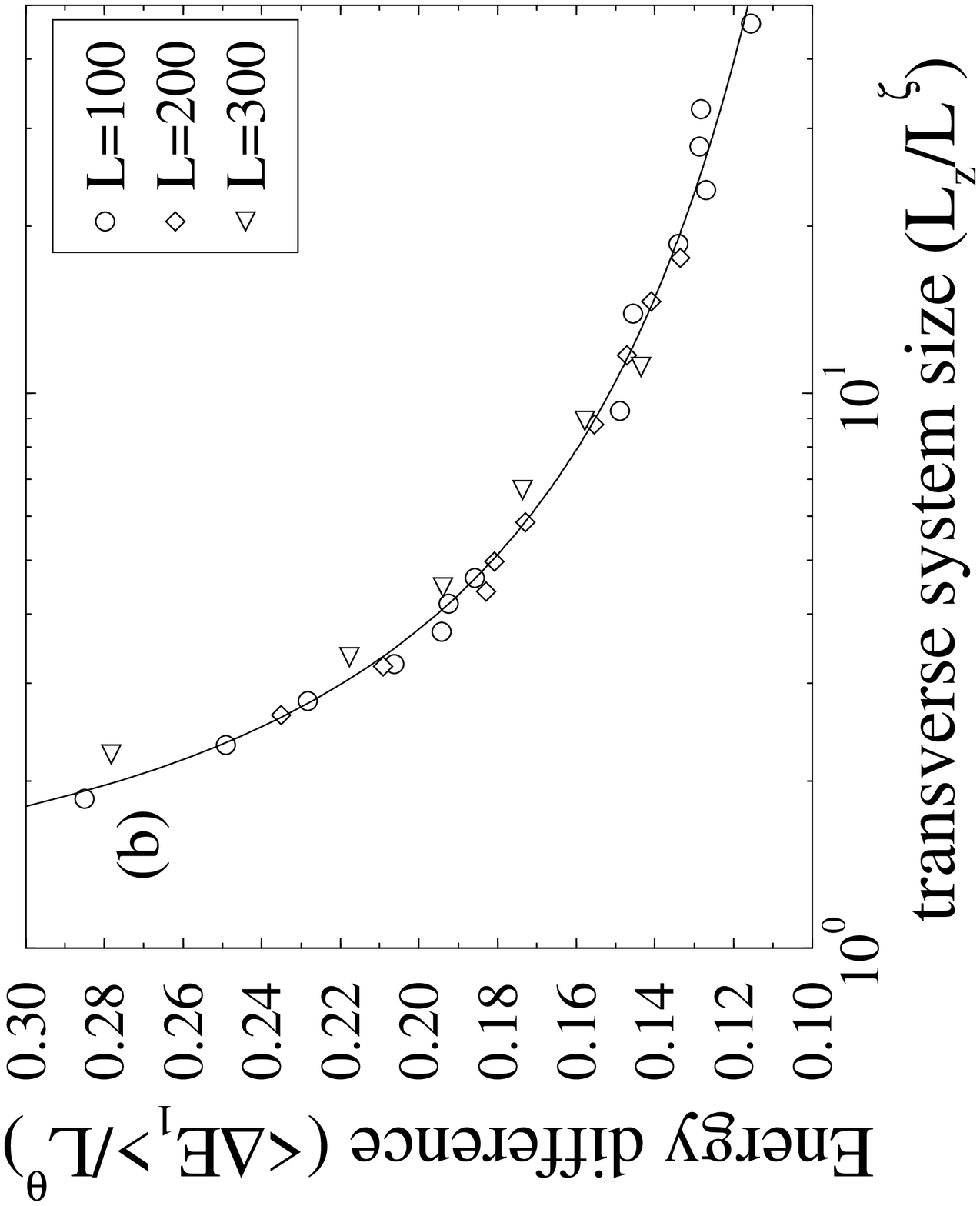}}
\caption{(a) The scaling function of the probability 
distribution $\hat{P}(\Delta E_1/ \langle \Delta E_1 \rangle)\times \langle 
\Delta E_1 \rangle$ for the energy gaps $\Delta E_1$ of 
the lowest two minima normalized by their disorder-average $\langle 
\Delta E_1 \rangle$ in a (10-base) semilog-scale for the system sizes 
$L \times L_z = L^2 =100^2$ and $200^2$. The inset shows the tails in the
natural-log-scale. The initial global minimum position $\bar{z}_0/L_z
\simeq \rm{const}$ for all $L$. The number of realizations 
$N=2 \times 10^4$ for both system sizes. The lines in the figure and
in the inset are $\exp(-x)$.  (b) The scaling function $f(y)$ of the
scaled disorder-average of the energy gap $\langle \Delta E_1 \rangle
/L^{\theta}$ as a function of scaled transverse system size
$L_z/L^{\zeta}$ for the system sizes $L=$100, 200, and $300$, each
with $\bar{z}_0/L_z \simeq \rm{const}$. $\theta =1/3$, $\zeta
=2/3$. The line $f(y) = 0.23 \ln(y)^{-1/2}$ is a guide to the eye. The
number of realizations ranges from $N=500$ for $L=300$, $L_z=500$ to
$N=2000$ for $L=200$, $L_z=600$. Only the non-overlapping interfaces
are considered.  The fraction of overlapping ones is vanishingly
small.}
\label{fig7}
\end{figure}

The probability distributions for the energy gaps of directed polymers
in system sizes $L^2 =100^2$ and $200^2$ are shown in
Fig.~\ref{fig7}(a). The distribution has a finite value at $\Delta E_1
=0$ and the tail has approximately a stretched exponential behavior,
$\exp(-ax^b)$, with an exponent $b \simeq 1.3$. In the figure there is
plotted as a comparison an exponential $\exp(-x)$ line, from which the
deviation of the probability distribution is more clearly seen in the
inset, where the distribution is in a natural-log scale. The finite
value at $\Delta E_1 =0$ is consistent with the weak replica symmetry
breaking picture.

To derive the scaling function for the $\Delta E_1$ it is expected
that in systems with height $L_z$ small enough to restrict the number
of minima $\Delta E_1$ mainly depends on the height of the system
$L_z$.  On the other hand, when $L_z$ is large enough, there are
enough valleys of which to choose the two minima, and one has $\Delta
E_1 \sim \Delta E \sim L^\theta$, hence
\begin{equation}
\langle \Delta E_1(L,L_z) \rangle \sim \left\{ \begin{array}{lll}
\tilde {f}(L_z),&\mbox{\hspace{5mm}}&L_z \ll L, \\
L^{\theta},& &L_z \gg L,
\end{array} \right.
\label{limits} 
\end{equation}
when the smaller parameter being varied.
Since it is assumed, that $L_z \sim L^\zeta$, a natural scaling form
based on these limiting behaviors is,
\begin{equation}
\langle \Delta E_1(L,L_z) \rangle\sim L^{\theta} f\left(\frac{L_z}
{L^{\zeta}}\right).
\label{scalingDE} 
\end{equation}
The argument $y = L_z/L^{\zeta}$ for the scaling function $f(y)$
is just a function of the number of the minima, i.e., $L_z/L^\zeta
\sim N_z$, and the scaling function has the form from Eq.~(\ref{gap}),
when $\eta=2$,
\begin{equation}
f(y) \sim [\ln y]^{-1/2}. 
\label{scalingf}
\label{Nscaling} 
\end{equation}
In Fig.~\ref{fig7}(b) we have plotted the scaling
function~(\ref{scalingf}) by collapsing $\langle \Delta E_1(L,L_z)
\rangle/L^\theta$ versus $L_z/L^{\zeta}$ for various $L$ and $L_z$,
and find a nice agreement confirming the scaling behavior
Eqs.~(\ref{scalingDE}), (\ref{scalingf}) as well as the analytic form
Eq.~(\ref{gap}) again assuming a Gaussian distribution ($\eta =2$).

Next we explore the first jump fields. Consider the relation of the
gap distribution and that of the jump fields given by
Eq.~(\ref{Ph1}). If we approximate $\hat{P}(\Delta E_1)$,
Eq.~(\ref{DE1_3}), with a uniform distribution, we get for the
probability distribution of the first jump field
\begin{equation}
P(h_1) = \exp \left[-\frac{N_z h_1}{2}\right] \frac{N_z}{h_1} 
\ln \left[ \frac{1- \frac{h_1}{N_z}} {1-h_1}\right] \sim \exp (-h_1).
\label{Ph}
\end{equation}
The form of $P(h_1)$ implies that it has a finite value at $h_1=0$,
which is again consistent with the weakly broken replica symmetry
picture~\cite{Parisi90} and also with the discussion in Section
\ref{argumentti}. We have also tried exponential and power-law type
of probability distributions for $\hat{P}(\Delta E_1)$ in
Eq.~(\ref{Ph1}), but all trials with negative exponents vanishes too
fast with $h_1$ compared to the numerical data, and all positive
exponents have behaviors with $P(h_1 \to 0) \to 0$ and diverge for
larger $h_1$.

\begin{figure}[f]
\centerline{\includegraphics[width=7cm,angle=-90]{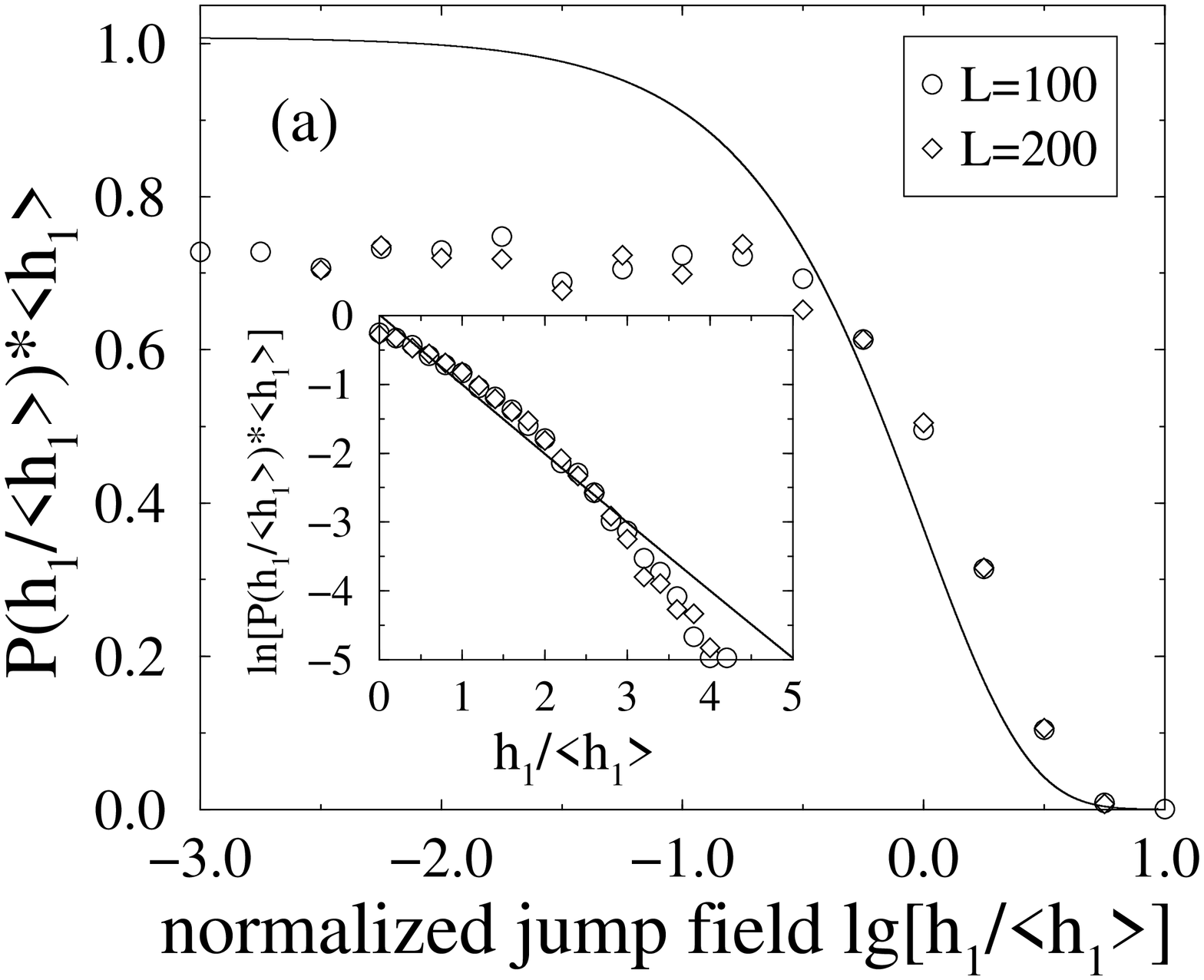}}
\centerline{\includegraphics[width=7cm,angle=-90]{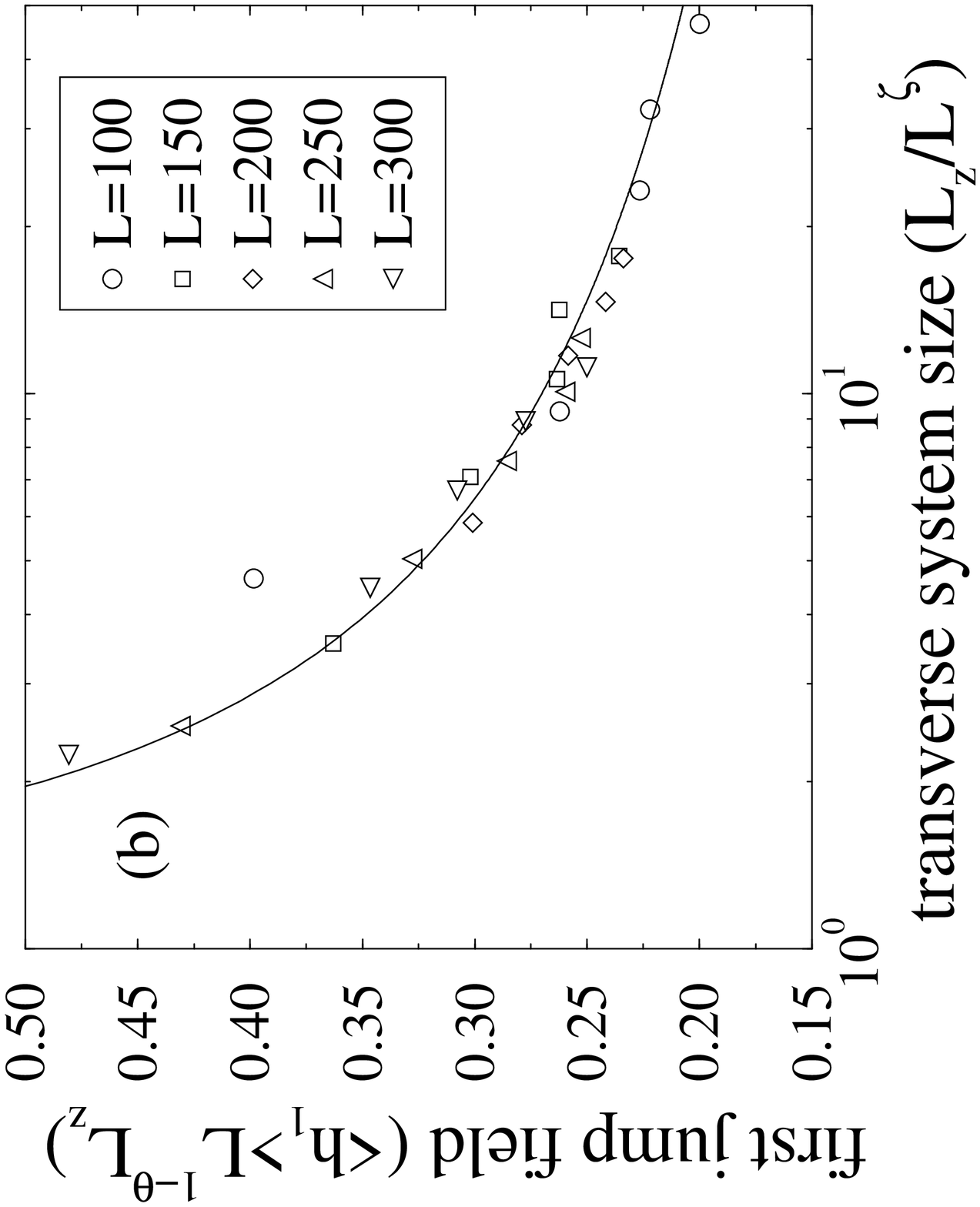}}
\caption{(a) The scaling function of the probability distribution $P(h_1/
\langle h_1 \rangle)\times \langle h_1 \rangle$ for the first jump
field values $h_1$ normalized by their disorder-average $\langle h_1
\rangle$ in a (10-base) semilog-scale for the system sizes $L \times
L_z = L^2 =100^2$ and $200^2$. The inset shows the tails in the
natural-log-scale. The initial global minimum position $\bar{z}_0/L_z
\simeq \rm{const}$ for all $L$. The number of realizations 
$N=2 \times 10^4$ for both system sizes. The line is the analytic
result from Eq.~(\ref{Ph1}) with a uniform distribution $\hat{P}(x)$ and
$N_z=20$. (b) The scaling function $f(y)$ of the scaled
disorder-average of the jump field $\langle h_1 \rangle L^{1-\theta}
L_z$ as a function of scaled transverse system size
$L_z/L^{\zeta}$ for the system sizes $L=$100, 150, 200, 250 and
$300$, each with $\bar{z}_0/L_z \simeq \rm{const}$. $\theta =1/3$,
$\zeta =2/3$.  The line $f(y) = 0.41 \ln(y)^{-1/2}$ is a guide to the
eye.  The number of realizations ranges from $N=500$ for $L=300$,
$L_z=500$ to $N=2600$ for $L=200$, $L_z=600$. Only the non-overlapping
jumps are considered, again.}
\label{fig8}
\end{figure}

In Fig. \ref{fig8}(a) we have plotted the probability distribution of
the first jump field for the system sizes $L^2 =100^2$ and $200^2$.
The probability distribution of the first jump field is similar to the
probability distribution of the gap energies. The analytic formula,
Eq.~(\ref{Ph}), is drawn as a line in the figure.  One clearly sees
that the line is a pure exponential, $\exp(-x)$, and the deviation of
the numerical data from the exponential is similar to
Fig.~\ref{fig7}(a) of the energy gap distributions.  Hence the shape
of the numerical first jump field distribution is approximately a
stretched exponential, $\exp(-ax^b)$, with an exponent $b \simeq
1.3$. Based on Figs.~\ref{fig7}(a) and \ref{fig8}(a) one sees, that
the probability distributions of the energy gaps and first jump fields
are the same. Thus with the correct $\hat{P}(\Delta E)$ one gets from
Eq.~(\ref{Ph1}) the same $P(h_1)=\hat{P}(\Delta E)$. This distribution
may be for small $h_1$ flat, but obviously starts to vanish for larger
$h_1$ since $P$ decays faster than exponentially.  Another reason can
be the fact that by discarding samples with the GS outside of a
predefined window we just constrain the number of valleys in the
sample so that the expectation value is the same in each one. $N_z$
can however fluctuate from sample to sample.  The most important
consequence is in any case that $\hat{P}(\Delta E_1=0)=P(h_1=0)
\not=0$. The finite size scaling and the normalization of the
probability distribution of $\Delta E_1$ give $\hat{P}(\Delta
E_1=0) \sim L^{-\theta} [\ln(L_z L^{-\zeta})]^{1/\eta}$.

In order to find the scaling relation for the first jump field $h_1$,
we make the Ansatz $\langle \Delta E_1 \rangle = \langle h_1 \rangle L
L_z$, since the field contributes to the manifold energy proportional
to $L^D$ ($D=1$) and $L_z \sim \langle \Delta z_1 \rangle$ is the
difference in the field contributions $hz$ to the energy at finite $h$
at different average valley heights $z_0$, $z_1$.  Hence
\begin{equation}
\langle h_1(L,L_z) \rangle L L_z \sim L^{\theta} 
f\left(\frac{L_z}{L^{\zeta}}\right),
\label{scalingh1} 
\end{equation}
where the scaling function $f(y)$ for the number of the minima $N_z
\sim L_z/L^\zeta \sim y$ has the same scaling function
Eq.~(\ref{Nscaling}).  Fig.~\ref{fig8}(b) shows the scaling
function~(\ref{scalingf}) with a collapse of $\langle h_1(L,L_z)
\rangle L^{1-\theta}L_z$ versus $L_z/L^{\zeta}$ for various $L$ and
$L_z$ with a good agreement, again. 

\begin{figure}[f]
\centerline{\includegraphics[width=7cm,angle=-90]{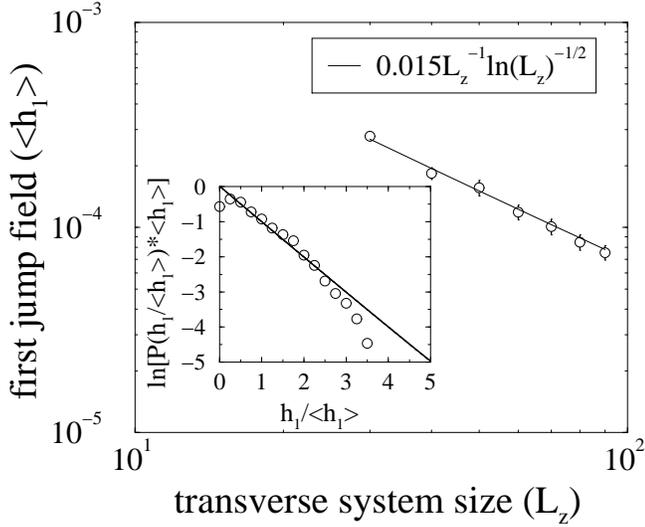}}
\caption{The disorder-average of the first jump field $\langle h_1
\rangle$ as a function of transverse system size $L_z$ for the 
$(2+1)$ dimensional interface. The disorder is random bond type, with
$\delta J_{ij}/J_0=1$, the number of realizations for each system
size $N \ge 100$, and the system size $L_x=L_y=50$. The initial 
global minimum position $\bar{z}_0/L_z \simeq  \rm{const}$. The line
$0.015 L_z^{-1} [\ln(L_z)]^{-1/2}$ is a guide to the eye. The inset shows 
the tail of the scaling function of the probability distribution $P(h_1/
\langle h_1 \rangle)\times \langle h_1 \rangle$ for the first jump
field values $h_1$ normalized by the disorder-average $\langle h_1
\rangle$ in a  natural-log-scale for the system size $L^3=50^3$.
The initial global minimum position $\bar{z}_0/L_z \simeq \rm{const}$.
The disorder is dilution type with $p=0.5$ and the first
non-overlapping jump is considered.  The number of realizations
$N>3000$. The line is the same as in the inset of Fig.~\ref{fig8}(a).}
\label{fig9}
\end{figure}

Next we move over to the $(2+1)$ dimensional manifolds. The inset of
Fig.~\ref{fig9} shows the tail of the distribution for the system size
$L^3 =50^3$ with dilution type of disorder and bond occupation
probability $p=0.5$. The first non-overlapping jumps are included in
the distribution, since due to the anomaly of the dilution disorder
(lots of small scale degeneracy), there are typically two adjustments
in the interface before the large jump. As a comparison, we plot again
the $\exp(-x)$ line, too. One sees that the deviation of the tail of
the distribution from the exponential behavior is similar to the
$(1+1)$ dimensional case. The finite scaling of the first jump field
in Fig.~\ref{fig9} is shown for interfaces of size $L_x
\times L_y = 50^2$ and systems of height $L_z = 30\--90$.  We have
fitted for a constant $L$ the formula (\ref{scalingh1}), i.e., $\langle
h_1(L_z) \rangle \sim L_z^{-1} [\ln(L_z)]^{-1/2}$ and it works within
error bars.

Generalizing the numerical results of $(1+1)$ and $(2+1)$ dimensional
calculations and the analytic arguments from the previous section to
arbitrary dimensions gives the behavior of $\langle h_1(L,L_z) \rangle 
\sim L^{\theta-D} L_z^{-1} [\ln(L_z/ L^{\zeta})]^{-1/2}$.  Since the
probability distribution has a finite value at $P(h_1 = 0)$ and
$\langle h_1(L,L_z) \rangle$ vanishes with increasing system size, one
obtains from the normalization factor at $P(h_1 = 0)$ for the scaling
of the susceptibility, Eq.~(\ref{Eqsuskis2})
\begin{equation}
\chi \sim L^{D-\theta} L_z [\ln(L_z/L^{\zeta})]^{1/2},
\label{Eqsuskis3}
\end{equation}
and in the isotropic limit, $L \propto L_z$, the total susceptibility 
$\chi_{tot} = L^d \chi$  becomes
\begin{equation}
\chi_{tot} \sim L^{2D+1-\theta} [(1-\zeta)\ln(L)]^{1/2}.
\label{Eqsuskistot}
\end{equation}
Thus the extreme statistics of energy landscapes shows up in the
susceptibility of random manifolds or domain walls in a form that
Eq.~(\ref{Eqsuskistot}) has a logarithmic multiplier and the scaling
behavior of the first jump field is due to the scaling of the energy
minima differences. This is in contrast to Shapir's result
\cite{Shapir91} and the logarithmic contribution is also missing from
our earlier paper \cite{Seppala00} where the valley energy scale was
taken to follow the standard $L^\theta$ assumption.  Therefore one can
conclude that the effect of extreme statistics is important in this
problem: since we look at the finer details of the landscape
``typical'' differences are not sufficient. Notice that for two
dimensional random field Ising magnets $\zeta \simeq 1$ at large
scales \cite{Seppala98} and the susceptibility does not
diverge~\cite{Aizenman}: the premise that $N_z > 1$ is broken in
that case. If the condition $N_z > 1$ is violated the extreme
statistics correction disappears.  For flat interfaces so that the
effective roughness exponent is zero ($\zeta_{eff}=0$), e.g. \{100\}
oriented RB interfaces with weak disorder and small system sizes
$L<L_c$, when
$\zeta=0$~\cite{Bouchaud92,Alava96,Raisanen98,unpublished2}, $\theta
=D/2$ and Eq.~(\ref{Eqsuskistot}) becomes $\chi_{tot} \sim
L^{2D+1-D/2} [\ln(L)]^{1/2} \sim L^4 [\ln(L)]^{1/2}$, when $D=2$.

\begin{figure}[f]
\centerline{\includegraphics[width=6cm,angle=-90]{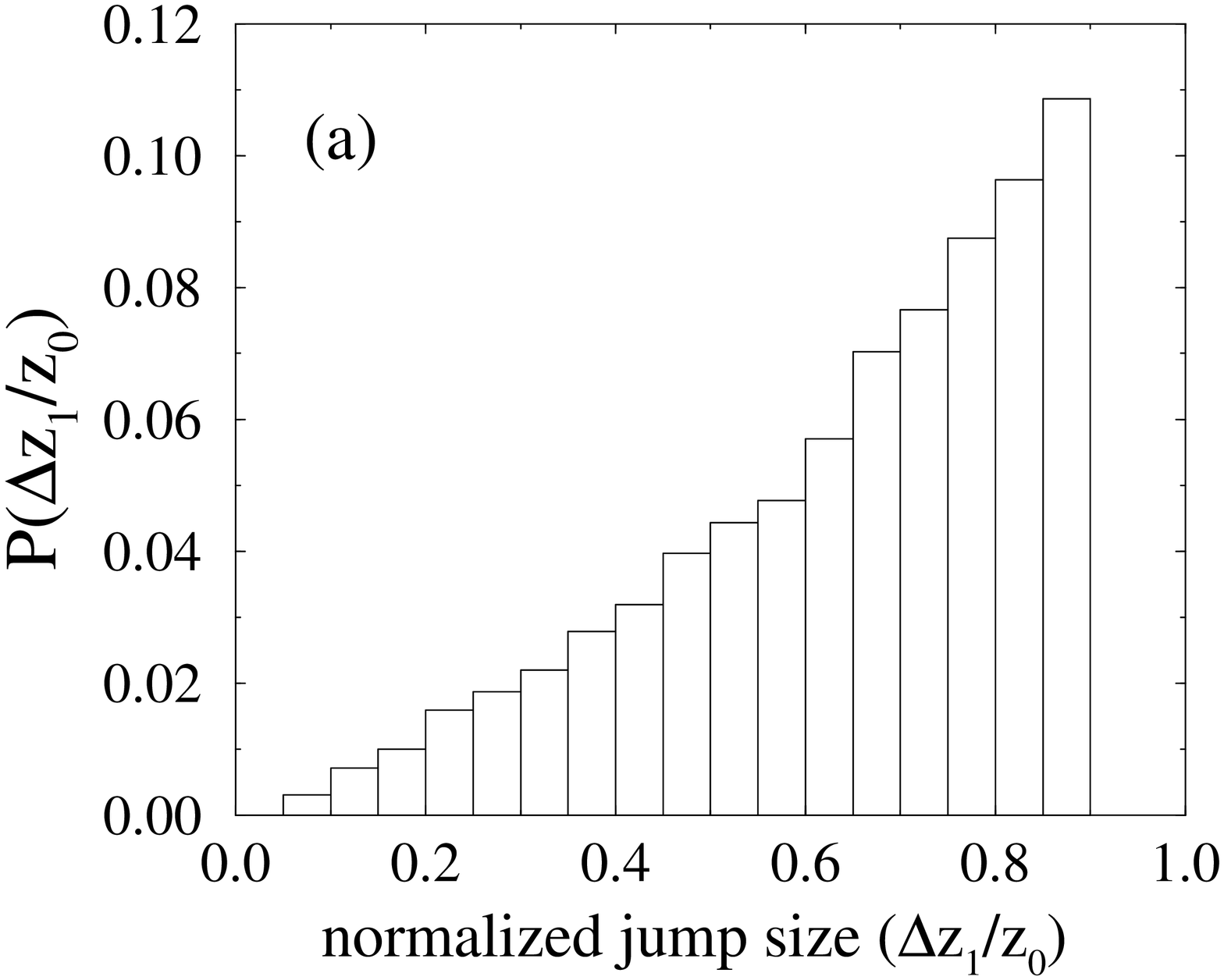}}
\centerline{\includegraphics[width=6cm,angle=-90]{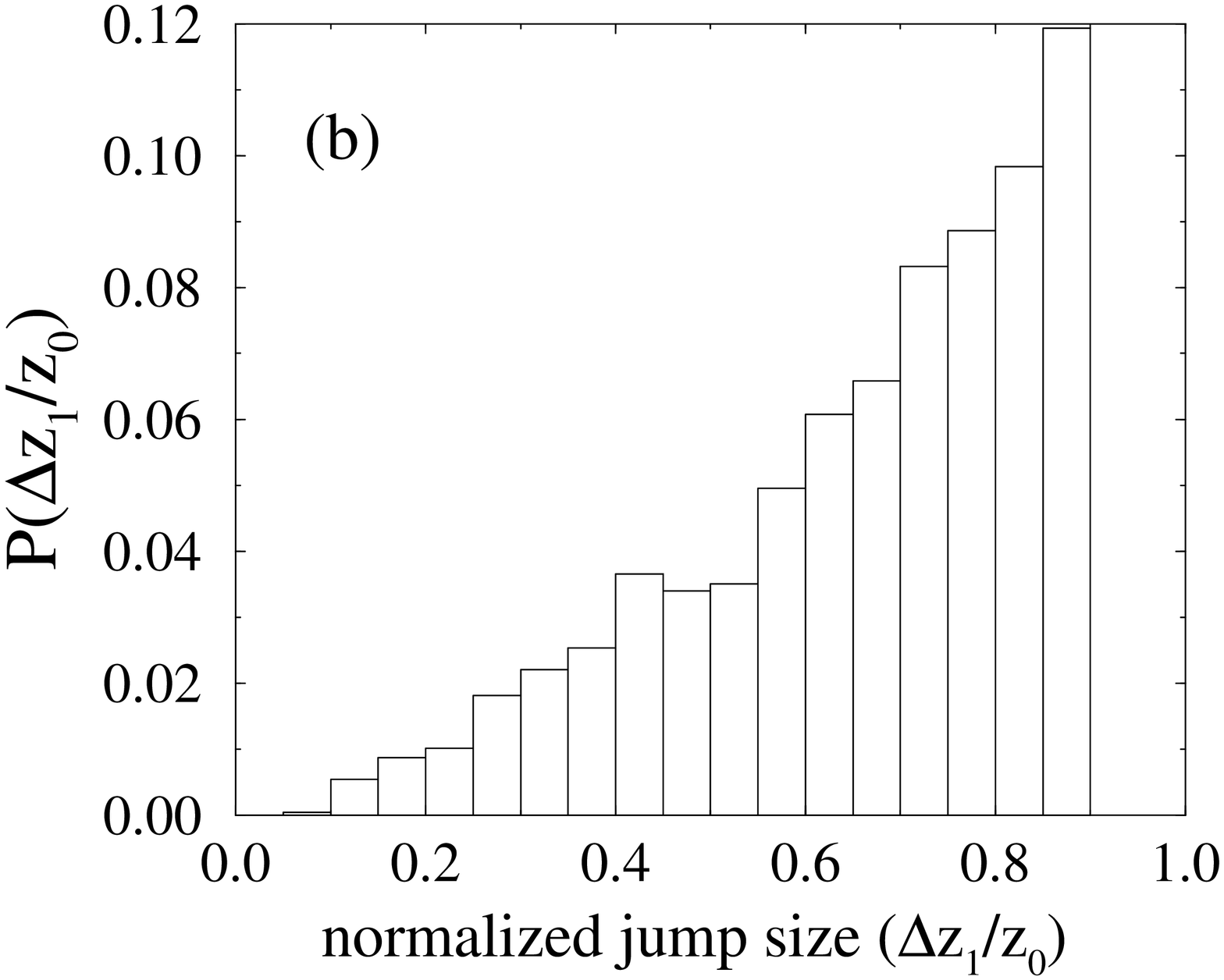}}
\centerline{\includegraphics[width=6cm,angle=-90]{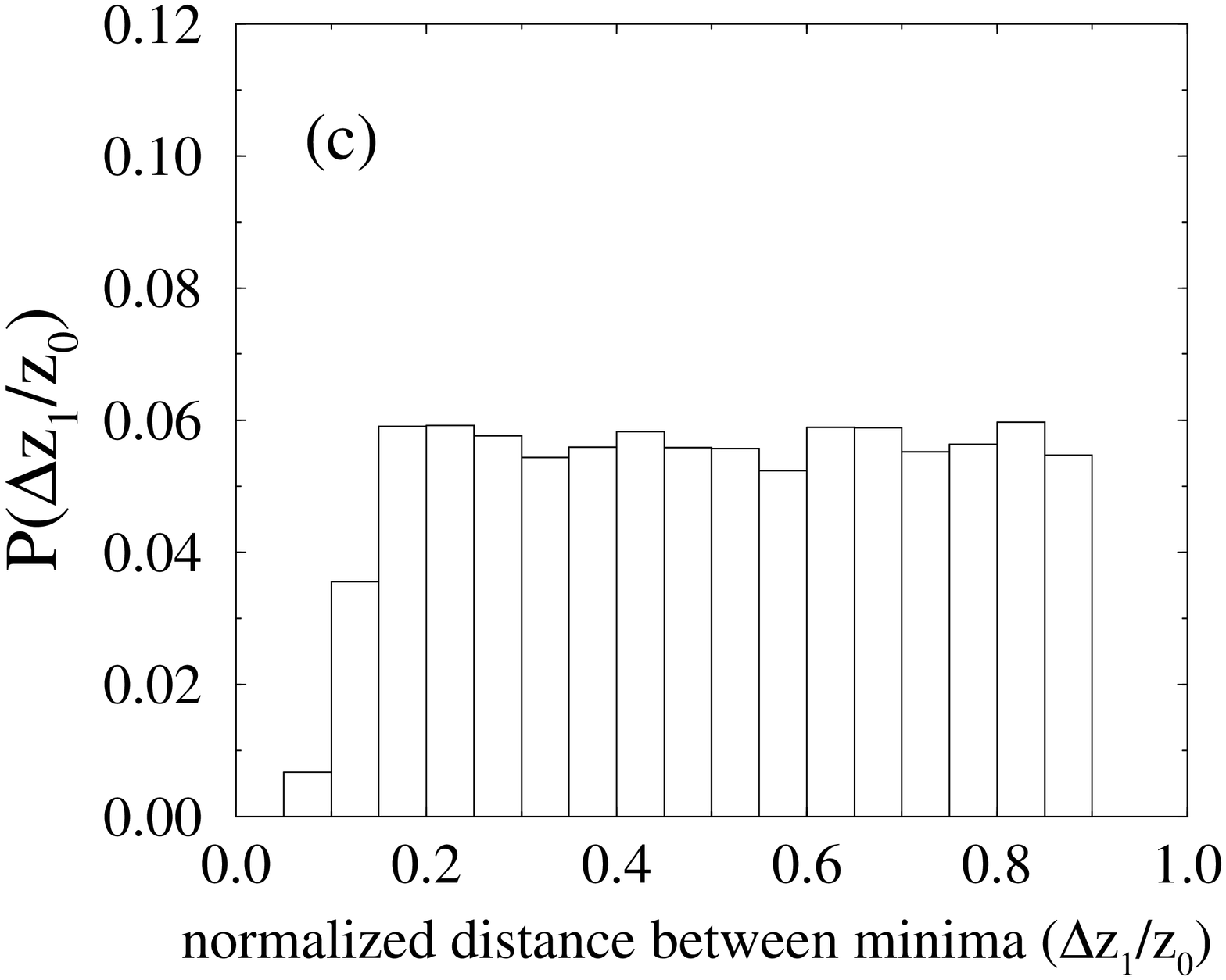}}
\caption{(a) The histogram $P(\Delta z_1/z_0)$ of the jump size
$\Delta z_1$ of the first jump normalized with the initial
position $z_0$ (the initial global minimum position $\bar{z}_0
\simeq \rm{const}$) for $(1+1)$ dimensional interface. The number 
of realizations $N=2 \times 10^4$, the system size $L^2=100^2$, and
the random configurations are the same as in Fig.~\ref{fig8}(a). 
(b) $P(\Delta z_1/z_0)$ for $(2+1)$ dimensional interface The number
of realizations $N \simeq 3000$ and the random configurations are the
same as for the data in Fig.~\ref{fig9} inset.  The system size
$L^3=50^3$ and disorder is dilution type with $p=0.5$.  The initial
global minimum position $\bar{z}_0 \simeq \rm{const}$.  (c) $P(\Delta
z_1/z_0)$ for $(1+1)$ dimensional interface without a field,
$L^2=100^2$, the number of realizations $N=2 \times 10^4$.  The random
configurations are the same as in Fig.~\ref{fig7}(a) and the initial
global minimum position $\bar{z}_0 \simeq \rm{const}$. Note in all
figures the repulsion induced behavior at the both ends: for values 
close to zero due to neglect of the overlapping interfaces and 
for values close to unity due to wall repulsion.}
\label{fig10}
\end{figure}

Finally we report the jump distance distributions of the first
jumps. Fig.~\ref{fig10}(a) shows $P(\Delta z_1/\overline{z}_0)$ of the
first jump with the field in $(1+1)$ dimensions. The distribution is a
superposition of two behaviors: the interface jumps to the lowest
minimum as such, and the external field favors the minima close to the
wall (remember the differences of $z_1$ and $z_{1^*}$ in
Fig.~\ref{fig5}). Since the non-overlapping cases are excluded and the
wall has a repulsive effect, there are cut-offs in the both ends of
the distributions. The shape of the distribution does not change with
the system size, which is consistent with the assumptions in
Eqs.~(\ref{Eqsuskis2}) and (\ref{scalingh1}), $\langle \Delta z_1
\rangle \sim L_z$. Fig.~\ref{fig10}(b) shows the same distribution of
$P(\Delta z_1/\overline{z}_0)$ with a field for $(2+1)$ dimensional
manifolds and it is clear that the shape of the distribution does not
depend on the dimension. In Fig.~\ref{fig10}(c) we have plotted
$P(\Delta z_1/\overline{z}_0)$ without a field, i.e., the distance of
the true lowest energy minima in $(1+1)$ dimensions. Now there is no
field, and thus the shape of the distribution is just a uniform one,
again consistent with the predictions made in Section~\ref{argumentti} in
Eq.~(\ref{P0prod}).

\section{Scaling of the first jumps: field and distance}\label{first_jump}

Following similar arguments as Eq.~(\ref{excitation}) in
Section~\ref{glassy} for an excitation, a mean field result for the
finite size scaling of the first jump field is next derived.  Let us
have an interface at an arbitrary height $z_0$ with an energy $E_0$
and an isotropic system $L \propto L_z$.  The energy gap between the
two lowest energy minima, Eq.~(\ref{gap}), scales as $\Delta E_1 \sim
L^{\theta} [ (1-\zeta) \ln (L)]^{-1/2}$, when the $z_0 \simeq L_z$.
However, when the interface is at an arbitrary height the number of
the available minima depends on the original position of the
interface, and we use only the dominating algebraic behavior so that
$\Delta E_1 \sim L^{\theta}$.

On the other hand the energy difference of elastic manifolds at
different heights due to the field contribution is $\Delta E \simeq h
\Delta z L^D$.  Assuming that $\langle \Delta z \rangle \sim L$ as in
the earlier arguments, the field contribution becomes $\Delta E \sim h
L^d$.  We expect that the first jump happens, when the gap equals the
field energy, and thus the first jump field has a scaling
\begin{equation}
\langle h_1 \rangle  \sim L^{\alpha} = L^{\theta-d}.
\label{MFh1}
\end{equation}

\begin{figure}[f]
\centerline{\includegraphics[width=7cm,angle=-90]{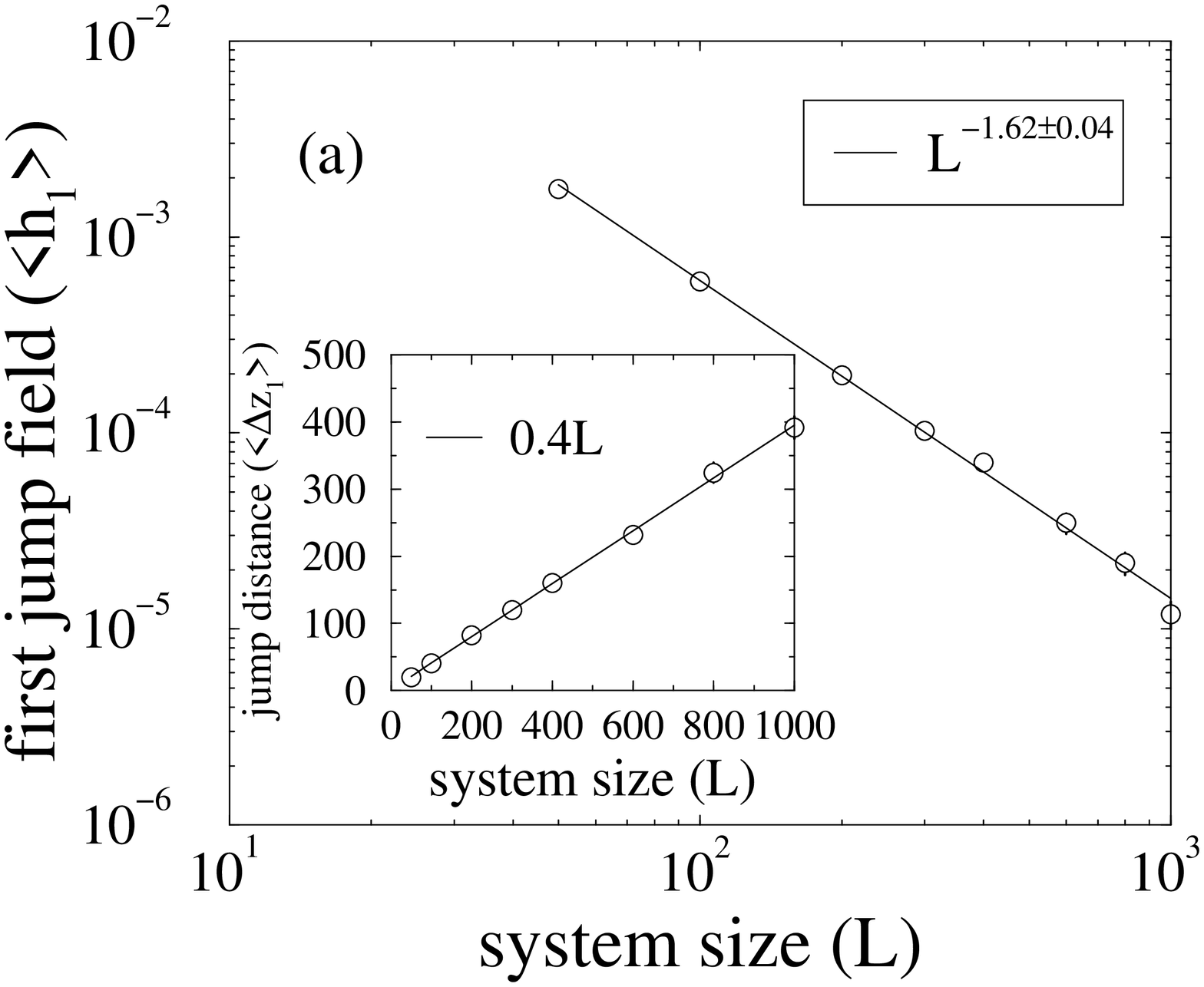}}
\centerline{\includegraphics[width=7cm,angle=-90]{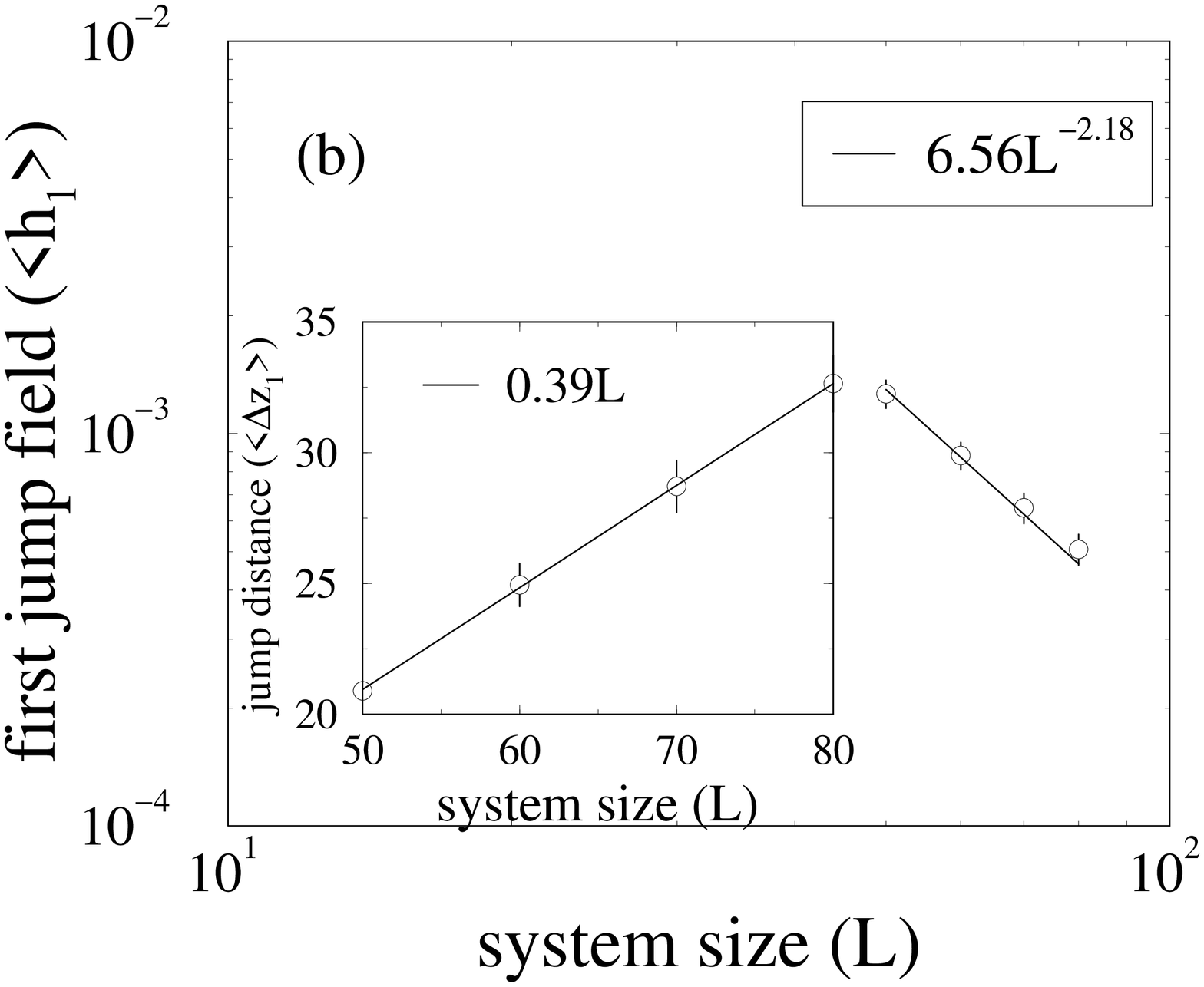}}
\caption{(a) Finite size scaling of the average first jump field $\langle
h_1 \rangle$ for DP's. The line is the least squares fit to data. The
scaling argument gives $\alpha = 5/3$.  The inset shows the average
jump distance $\langle \Delta z_1 \rangle$ at the corresponding field
$h_1$ with a linear fit to data.  $\langle \, \rangle$ is the
disorder-average over $N=1000$ realizations for the system sizes $L
\times L_z =L^2 = 50^2$ and $100^2$, $N=500$ for $L^2 =
200^2\--400^2$, and $N=200$ for $L^2 = 600^2\--1000^2$.  The disorder
is of random bond type. No overlaps included. (b) Finite size scaling
of the average first jump field $\langle h_1 \rangle$ for $(2+1)$
dimensional manifold. The line is a guide to the eye with a slope
$\alpha \simeq 2.18$. The inset shows the average jump distance $\langle
\Delta z_1 \rangle$ at the corresponding field $h_1$ with a linear fit
to data. $\langle \, \rangle$ is the disorder-average over $N \simeq
250$ realizations for each system size. The disorder is dilution type
with $p=0.5$ and the first non-overlapping jump is considered 
(usually two small-scale adjustments take place before the jump).}
\label{fig11}
\end{figure}

In Figs.~\ref{fig11}(a) and \ref{fig11}(b) we have plotted the average
first jump field $\langle h_1 \rangle$ in isotropic systems for
$(1+1)$ and $(2+1)$ dimensional manifolds, respectively.  The data is
taken from non-overlapping jumps, to minimize finite size effects
since they are smaller than with the overlapping jumps
included. However, the fraction of overlapping jumps is small as seen
in Fig.~\ref{fig4} for $(1+1)$ dimensional case. The data in
Figs.~\ref{fig11} confirms within error bars the expected behavior of
$\langle h_1 \rangle \sim L^{-5/3}$ and $\langle h_1
\rangle \sim L^{-2.18}$ for $(1+1)$ and $(2+1)$ dimensional manifolds,
respectively. The logarithmic term $[\ln (N_z)]^{-1/2}$ might
contribute a downward trend since it decreases with the number of the
available minima $N_z$, here disorder averaged due to the arbitrary GS
location. This is however not noticeable in the data: the finite size
effects in the calculated data are towards smaller absolute value of
the exponent. In the insets of Figs.~\ref{fig11}(a) and (b) the linear
scaling of jump sizes $\langle \Delta z_1 \rangle \sim L$ is
confirmed, which was assumed in Eqs.~(\ref{Eqsuskis2}) and
(\ref{scalingh1}).

\begin{figure}[f]
\centerline{\includegraphics[width=7cm,angle=-90]{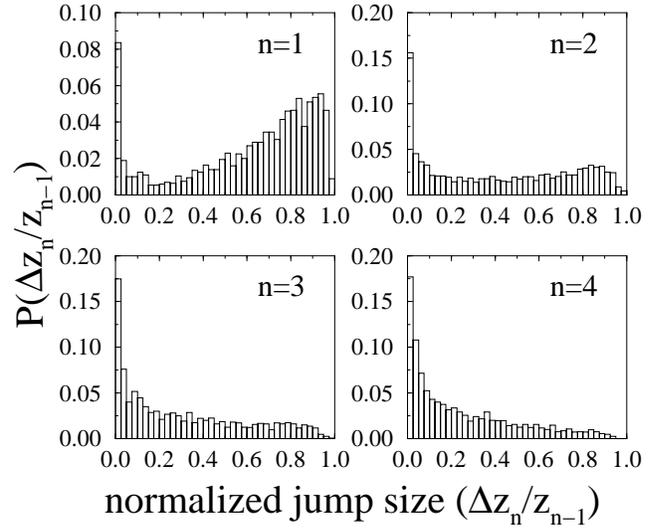}}
\caption{Evolution of the histogram of the jump sizes 
$\Delta z_n/z_{n-1}$ normalized with its previous position for 
$n=1,2,3,4$. The system size is $L^2 = 100^2$,  the number of 
realizations $N=2000$. The overlapping jumps are included.}
\label{fig12}
\end{figure}

The evolution of the jump size distribution for succeeding jumps is
demonstrated in Fig.~\ref{fig12}. As a comparison for Fig.~\ref{fig10}
where only the non-overlapping jumps were considered, here we have
shown the overlapping ones, too. It is seen as a peak near $\Delta
z_{n}/ \overline{z}_{n-1} = 0$. The first jumps have clearly the most
weight in the large jump end of the distribution, but for the
following jumps the whole distribution shifts towards $\Delta
z_{n}/\overline{z}_{n-1} = 0$.  After a small number of jumps the
interface is already in the minimum closest to the wall and then the
random-bulk wetting behavior takes over. In order to estimate the
number of the jumps the manifold does before binding to the wall, it
is assumed that after a jump the system looks statistically the same
as before it. From Fig.~\ref{fig10}(a) we infer that the probability
distribution for the jump size has a form $P(\Delta z) \sim \Delta
z^2$ (note that we did not attempt to compute this from
the analytical valley arguments) and that is then taken to
be the same for all (large) jumps $\Delta z_n$.  To calculate the
average jump size of jumps which {\it do not} jump to the closest
valley to the wall we have
\begin{eqnarray}
\langle \Delta z_n \rangle = \int_0^{z_{n-1}-A_1w} B_{n-1} \Delta z 
(\Delta z)^2 \, d(\Delta z) \nonumber \\
= \frac{3(z_{n-1}-A_1w)^4}{4z_{n-1}^3},
\label{Eqjumpsize}
\end{eqnarray} 
where the upper bound of the integral is to neglect the cases that
jump to the closest valley to the wall, $A_1$ is the prefactor to
multiply the roughness value to get the valley width, and $B_{n-1} =
3/z_{n-1}^3$ normalizes the probability distribution. Using 
$\langle z_n \rangle = \langle z_{n-1} \rangle - \langle \Delta z_n 
\rangle$ we get the next position of the interface. In order to 
calculate the probability of an interface to jump with the first jump
to the closest valley to the wall, $p_1$, we use the same probability
distribution as in Eq.~(\ref{Eqjumpsize}),
\begin{equation}
p_1 = \int_{z_0-A_1w}^{z_0} B_0 (\Delta z)^2 \, d(\Delta z) =
1-\left[1-\frac{A_1w}{z_0}\right]^3 = {\tilde p}_1,
\label{jumpp1}
\end{equation} 
and similarly for the probability of an interface to jump with
the second jump to the closest valley to the wall
\begin{eqnarray}
p_2 = (1-{\tilde p}_1) {\tilde p}_2 = (1-{\tilde p}_1)
\int_{z_1-A_1w}^{z_1} B_1 (\Delta z)^2 \, d(\Delta z) \nonumber \\ 
= \left[1-\frac{A_1w}{z_0}\right]^3 \left\{1-\left[1-\frac{A_1w}{z_1
}\right]^3 \right\}
\label{jumpp2}
\end{eqnarray} 
Due to the hierarchy we finally get for the $n^{th}$ jump
\begin{equation}
p_n = {\tilde p}_n \prod_{k=1}^{n-1} (1-{\tilde p}_k) =
\left[1-\left(1-q_{n}\right)^3 \right] \prod_{k=1}^{n-1} (1-q_k)^3,
\label{jumppn}
\end{equation}
where
\begin{equation}
q_n = \frac{A_1w}{z_{n-1}}.
\label{qfrac}
\end{equation}
To get an estimate for the number of jumps, we write
\begin{equation}
\langle {\mathcal N} \rangle = \sum_{n=1}^{\infty} np_n.
\label{n_estim}
\end{equation}
This can be evaluated, but only approximately since among others the
estimate of Eq.~(\ref{jumppn}) breaks down beyond $n$ small.  Using
$A_1w \simeq AL^\zeta= 50$, $L = 1000$, which is the case with $A_1
\simeq 6.5$, (see the numerics in Section~\ref{results}), and taking
$\overline{z}_0 =1000$, we get $\langle {\mathcal N} \rangle \simeq
3$. Eq.~(\ref{Eqjumpsize}) gives for the first jump size, with the
above numerical values, $\langle \Delta z_1 \rangle \simeq 0.6 
\overline{z}_0$, which is not far from the behavior in the inset 
of Fig.~\ref{fig11}(a), where $\langle \Delta z_1 \rangle \sim 0.4 
L_z \sim 0.8 \overline{z}_0$, since $\langle \overline{z}_0 \rangle  
\sim L_z/2$. 

\section{Application of a non-zero field: random-bulk wetting}\label{wetting}

\begin{figure}[f]
\centerline{\includegraphics[width=7cm,angle=-90]{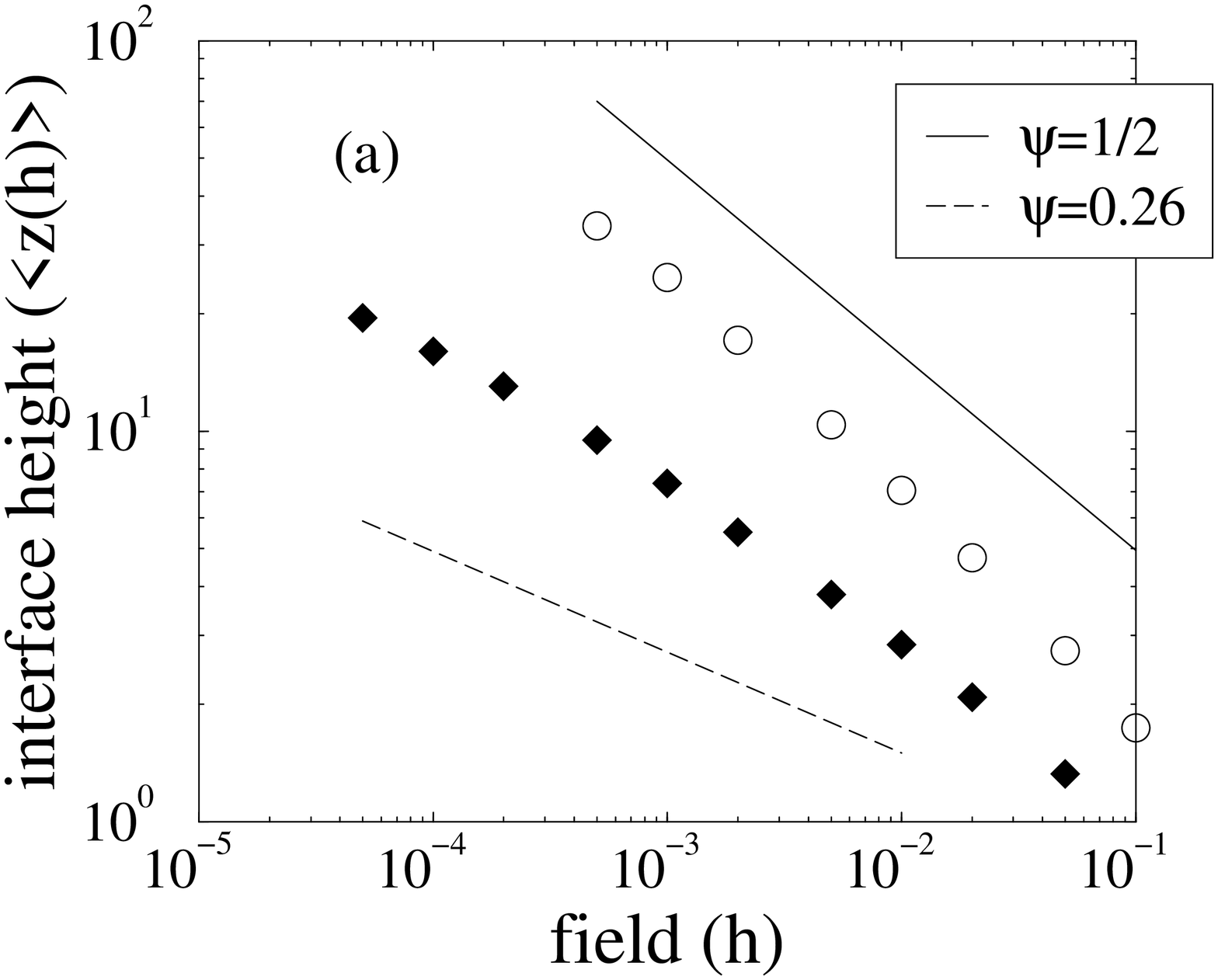}}
\centerline{\includegraphics[width=7cm,angle=-90]{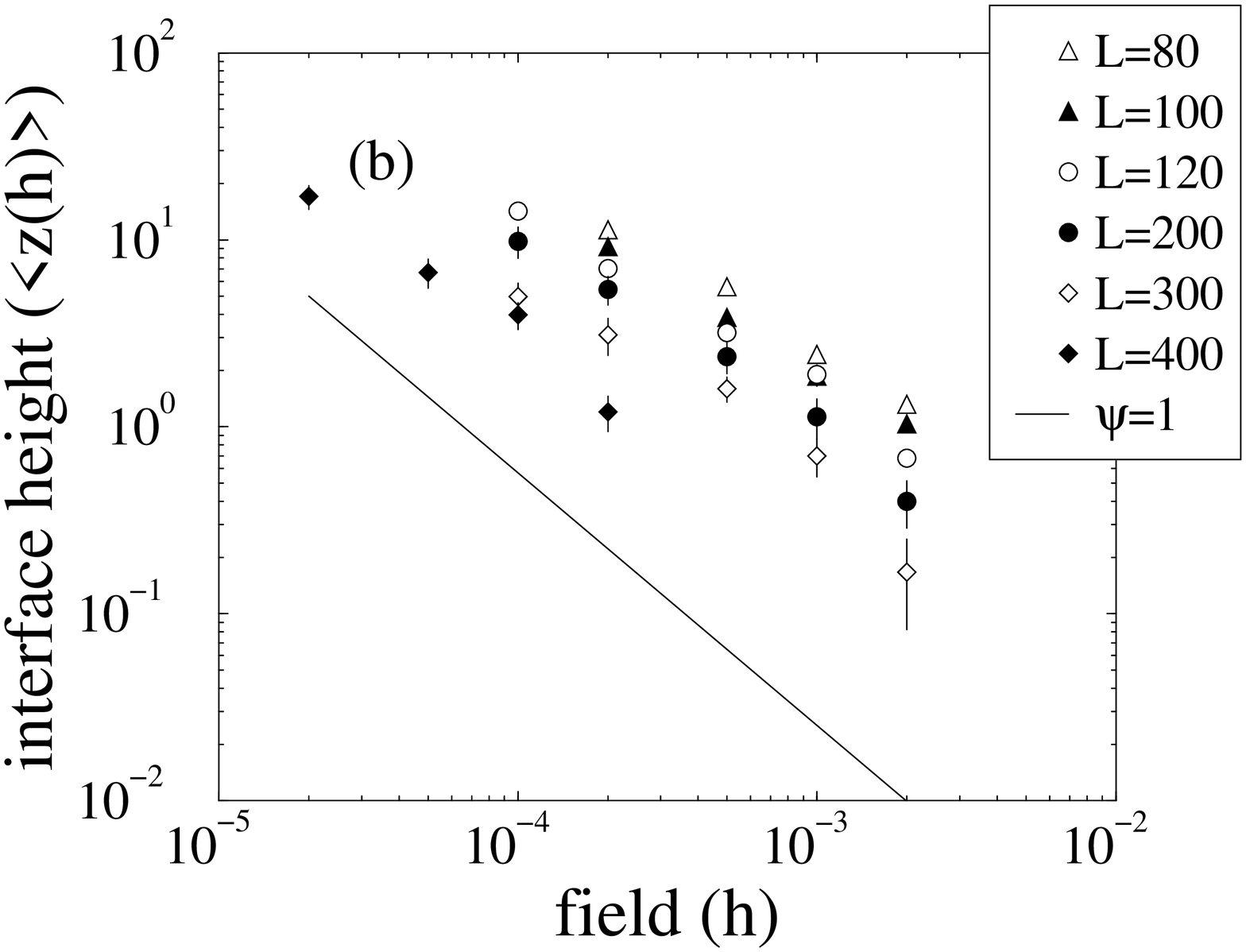}}
\caption{(a) The average interface mean height $\langle \bar{z}(h) 
\rangle$, i.e., the thickness of wetting layer, as a function of 
the external field $h$ for one dimensional directed polymer, open
circles. The system size is $L=3000$, $L_z=100$ and the number of
realizations $N=1000$. The disorder is dilution type with $p=0.55$.
The filled diamonds denote $(2+1)$ dimensional interfaces. The system
size is $L_x=L_y=300$, $L_z=50$ and the number of realizations $N=30$.
The disorder is dilution type with $p=0.30$. The solid line is a guide
to the eye with a slope $\psi=1/2$ and the dashed line has a slope
$\psi=0.26$. (b) $\langle \bar{z}(h) \rangle$ vs. $h$ for various
system sizes $L_x=L_y=L=80,100,120,200,300$, and $400$ in the flat
regime. The disorder is dilution type with $p=0.95$. $L_z=50$ for each
system.  The number of random configurations is $N=100\--200$ for
$L=80\--120$, and $N=30$ for $L=200\--400$. The line is a guide to the
eye with a slope $\psi=1$.}
\label{fig13}
\end{figure}

In order to see the wetting behavior one just studies the effect of
a large external field on the average interface to wall
distance. Notice that the low-field physics discussed extensively
above can be considered as the eventual outcome in any sample with
$L_z > L^\zeta$ so that more than one valley is available. This means
that when the field is decreased from a large field value, the
interface finally jumps into the bulk. Fig.~\ref{fig13}(a) we show the
average mean height $\langle \overline{z}(h) \rangle$ versus the field
$h$ for $(1+1)$ and $(2+1)$ dimensional manifolds. The system sizes
are $L_z \times L = 100 \times 3000$ and $L_z \times L^2 =50\times
300^2$.  To maximize the prefactor, $A_2$, in the scaling of
roughness, $w \sim A_2L^\zeta$, and hence the width of the minimum
energy valley, the disorder has been chosen to be strong, i.e.,
dilution type of disorder with small $p$, but above the
bond-percolation threshold. However, there are still some deviations
in the form of greater exponents than the expected from
Eq.~(\ref{psi}), which gives the values $\psi =1/2$ and $\psi \simeq
0.26$ in $(1+1)$ and $(2+1)$ dimensions from $\zeta_{(1+1)}=2/3$ and
$\zeta_{(2+1)}=0.41\pm 0.01$, respectively. We found that the trend is
nevertheless clear, with greater $L$ and fixed $L_z$ the exponents
become closer to the expected one. The effective exponent
$\psi_{eff}(L)$ can be used to extract the asymptotic, $L$-independent
exponent in particular in $(2+1)$ dimensions since this case is
hampered most by finite-size effects. $\psi_{eff}(L)$ as a function of
$1/L$ indicates that the asymptotic value is indeed $0.26 \pm 0.02$
and that the system sizes at which $\psi_{eff}(L)$ approaches that are
of the order of $L = 10^4$. That is, only with $10^{10}$ sites it
becomes possible to reach the asymptotic regime.  When one has $L
\propto L_z$ calculating the average mean height $\langle
\overline{z}(h) \rangle$ with a fixed field $h$ is nothing but
averaging over the jumped and not jumped interfaces together with
their location, see Fig.~\ref{fig2}(a).

In Fig.~\ref{fig13}(b) we have plotted as a comparison the average
mean height $\langle \overline{z}(h) \rangle$ versus the field $h$ for
$(2+1)$ dimensional manifolds with a weak disorder. In this case the
{\it weak} means for system sizes used, that the roughness of the
manifolds are not in the asymptotic roughness limit yet,
$L<L_c$~\cite{Bouchaud92,Alava96,Raisanen98,unpublished2}.  The system
sizes used are $L_z \times L^2 =50\times 80^2\--400^2$ and the
behavior is simple: either the interface stays in its original
position or jumps directly to the wall. Taking into account the jumped
and original interfaces as $\langle z(h) \rangle = \langle z
[1-P(\overline{z}_0,h)] \overline{z}_0 + P(\overline{z}_0,h) \times 0
\rangle = \langle [1-P(\overline{z}_0,h)] \overline{z}_0 \rangle =
\int_0^{L_z} [1-P\{\overline{z}_0(h)\}] \overline{z}_0 \,
d\overline{z}_0$, gives the behavior $\langle z(h) \rangle \sim
h^{-1}$, i.e., the exponent $\psi \simeq 1$, if $P\{\overline{z}_0(h)\}
\sim h^{-1}$.  The larger manifolds jump faster to the wall, i.e.,
they feel the perturbation earlier, since Eq.~(\ref{scalingh1}) for
flat interfaces ($\zeta =0$) becomes $\langle h_1(L,L_z) \rangle \sim
L^{\theta-D} L_z^{-1} [\ln(L_z)]^{-1/2}$. With fixed $L_z$ and $\theta
=D/2$ from Poissonian statistics, we get in $D=2$ $\langle h_1(L)
\rangle \sim L^{-1}$ consistent with the numerical data.  This leads
to the behavior of the wetting scaling, $\langle \bar{z}(h) \rangle
\sim c(L) h^{-\psi}$ where $c(L) \simeq L^{-1}$ and $\psi=1$. The
finite size scaling of the prefactor indicates that at large $L$-limit
with fixed $L_z$ the flat interfaces are immediately at the wall, and
thus the systems are non-wet.  This implies that there is an
interesting cross-over around $L_c$ between such a ``dry'' regime and
the bulk wetting that takes over for still larger $L$. In $D>2$ this
kind of behavior is relevant even in the thermodynamic limit, if the
disorder is weak due to the presence of a bulk roughening transition
for bond disorder.

\section{Discussion}\label{disc}

\subsection{Finite temperature behavior}\label{creep}

The movement of the elastic manifolds in random media at low
temperatures, when an applied force is much below the depinning
threshold $F_c$, is characterized as creep.  The dynamics is
controlled by thermally activated jumps over pinning energy barriers,
which separate the metastable states.  D.\ Fisher and Huse~\cite{Fish91}
showed that for a DP at finite temperature $T>0$ the fluctuations of
the entropy $(\Delta S)^2$ and the internal energy $(\Delta
E_{int})^2$ scale linearly with the length of the polymer and cancel
each other. Hence there are only the fluctuations of the free energy
$(\Delta F)^2$, which scale with the zero temperature energy
fluctuation exponent $2\theta=2/3$.  Since the free energy is the one
which should be minimized at finite temperature, it is the one which
defines statistically the shape of the energy landscape, although the
energy valleys and minima need not to have exactly the same real space
structure as at $T=0$. Thus the $\theta=1/3$ exponent should define
the energy gaps also when $T>0$, expect in the cases there is a
critical temperature $T_c$.  Hence, our derivation of the
susceptibility and also the first order character in the
reorganization of valleys should be relevant also at $T>0$.

\subsection{First arrival times in nonlinear surface growth}\label{secKPZ}

\begin{figure}[f]
\centerline{\includegraphics[width=9cm]{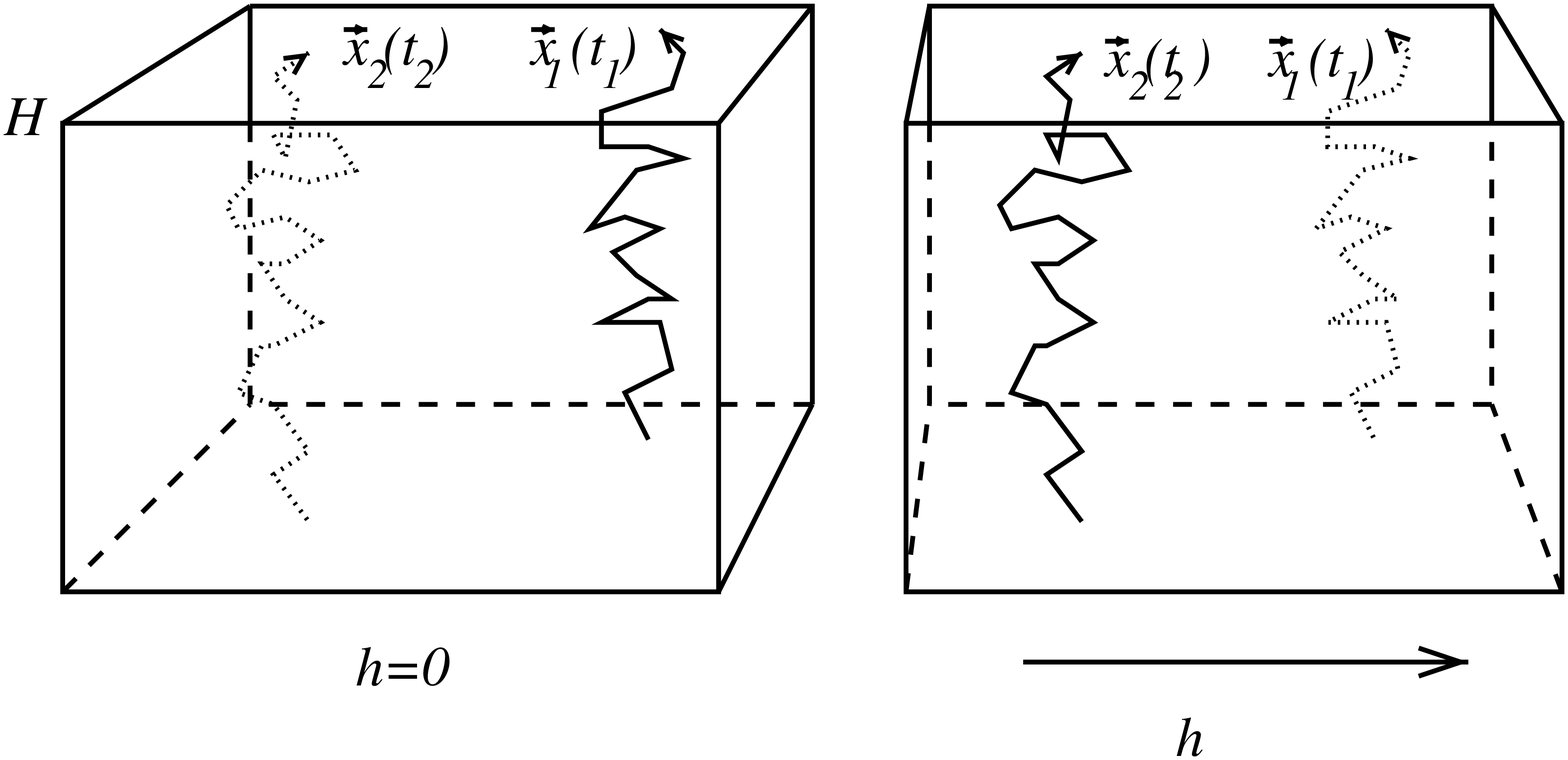}}
\caption{The relation between directed polymers and growing
interfaces.  Two directed polymers in independent valleys equal the
fastest arrival time $t_1$ at $\vec{x}_1$, solid line, and the second
fastest at $\vec{x}_2$ with time $t_2$, dotted line, of a KPZ
interface to a fixed height $H$, when the external field $h=0$. In the
right hand side figure an external field is added, which increases the
growth time depending on the position in the direction of the arrow
and the polymer at $\vec{x}_2$ becomes the one corresponding
to the fastest time to reach $H$.}
\label{fig14}
\end{figure}

The $(1+n)$ dimensional directed polymers map, in the continuum limit,
to the KPZ~\cite{KPZ,Barab} equation by associating the minimum energy
of a DP-configuration with the minimum {\it arrival time} $t_1 \equiv
E_0$ of a KPZ-surface to height $H$. The connection is illustrated in
Fig.~\ref{fig14}.  The minimal path of the DP with the end point ${\bf
x}_1(t_1)$ equals the path by which the interface reaches $H$, at
location ${\bf x}_1$ and at time $t_1$. Thus $t_1$ attains a
logarithmic correction, from Eq.~(\ref{typicalene}), of size
$-H^{\beta} \{\ln(L/H^{1/z}) \}^{1/\eta}$, where $L$ is the linear
size of the system, $\beta = \theta > 0$ and $z=1/\zeta$ are now the
roughening exponent and dynamical exponent of the KPZ universality
class and the values of $\theta = 2\zeta-1$ and $\zeta >1/2$ depend on
$n$.  Notice that if there is a upper critical dimension $n_c$ in the
KPZ growth, then the logarithmic correction is not there anymore, and
$\theta=0$, $\zeta=1/2$, i.e., a random walk ensues. Consider now the
second smallest arrival time $t_2$. In the language of directed
polymers, if the path ${\bf x}_2 (t')$ corresponding to $t_2$ is
independent of ${\bf x}_1 (t')$ that corresponds to $t_1$ the time and
the path are found inside a separate, independent valley.  The {\it
difference} $\Delta t = t_2 - t_1$ then equals $\Delta E_1$, and
likewise obeys extreme statistics, so that $\Delta t \sim H^\beta
[\ln (L/H^{1/z})]^{-(\eta-1)/\eta}$.  For growing surfaces this limit
is the {\it early stages} of growth, in which the correlation length
$\xi \ll L$, and therefore the arrival times, or directed polymer
energies, are independent. On the other hand if we disturb the growing
process such a way that it depends on the ${\bf x}$, e.g.  linearly
with a factor $h$~\cite{spohn}, the polymer ending at ${\bf x}_2$
becomes faster if $h>h_1$, see right hand side of
Fig.~\ref{fig14}. Similarly now the factor $h_1$ has a scaling
behavior from Eq.~(\ref{scalingh1}), $h_1
\sim H^{\beta-1} L^{-1} [\ln (L/H^{1/z})]^{-(\eta-1)/\eta}$.

\section{Conclusions}\label{concl}

To conclude we have studied the $(1+1)$ and $(2+1)$ dimensional
elastic manifolds at zero temperature, when an external field is
applied. We have demonstrated that the response of manifolds shows a
first order character (``jump'') in the sense that the manifolds change
their configuration in the large system size limit completely. This
persists in a finite system over a small number of such jumps. The
distance that the center-of-mass moves is extensive. The whole picture
is based on a level-crossing between two low-energy valleys in the
energy landscape.  Averaging over the total magnetization of such
random magnets with a domain wall or over the positions of manifolds
when an external field is applied becomes dependent on whether the DW
or manifold has jumped or not. This leads to the problem of
self-averaging in random systems. Here a disorder average smooths
over the ``coexistence'' between systems that are not affected by a
finite field $h$ and those that have responded.

In order to study the susceptibility of the DW in random media, one
has to take into account the probability distribution of the
sample-dependent field associated with the global change of the
configuration, and take the limit of vanishing fields. This
probability distribution has a finite density at $h=0$. The finite
size scaling of the density is dependent on the finite size scaling of
the number of the low lying nearly degenerate energy minima in the
system.  The scaling of the number of the energy minima leads to a
logarithmic factor in the susceptibility, and can be accounted for by
using extreme statistics.  Such effects are difficult to study by
usual field theoretical means, since one has to have access to the
whole probability distribution and not only the few first moments
thereof. Notice that the crucial difference to much previous work is
the simple fact that we allow for multiple minima in the energy
landscape, which is often most excluded by the boundary conditions
applied to the problem.  Although the derivations and the numerical
calculations done here have concentrated on random bond type of
randomness similar behavior should be seen in random field cases, too.

The discrete character in the movement of elastic manifolds with an
external field results also in that the continuum theory for wetting
in random systems works only in slab geometries, where there is room
only for a single valley, or in the large external field limit, when
the interface is close to the wall. On the other hand the flat
interfaces are shown to jump directly to the wall, i.e., to be non-wet.
It would be interesting to see if the dynamics of the manifolds at
finite temperature reflects the first order character seen here at
$T=0$. Through the connection of (1+1) dimensional DW's to the
directed polymers of the KPZ surface growth, we have shown that the
logarithmic factor is also present in the statistics of growth times
in nonlinear surface growth.

\begin{acknowledgement}

The authors would like to thank the Academy of Finland's Centre of
Excellence Programme for financial support and the Center for
Scientific Computing, Espoo, Finland for computing resources. Phil
Duxbury and Simone Artz are acknowledged for many valuable
discussions.

\end{acknowledgement}


\begin{thebibliography}{}

\bibitem{Fis86} 
D.\ Fisher, 
Phys.\ Rev.\ Lett.\ \textbf{56} (1986), 1964.

\bibitem{Emig99} 
T.\ Emig and T.\ Nattermann, 
Eur. Phys.\ J.\ B, \textbf{8} (1999), 525.

\bibitem{KPZ} 
M.\ Kardar, G.\ Parisi, and Y.-C.\ Zhang,
Phys.\ Rev.\ Lett.\ \textbf{56} (1986), 889.

\bibitem{HaH95} 
T.\ Halpin-Healy and Y.-C.\ Zhang,
Phys. Rep. \textbf{254} (1995), 215.

\bibitem{Las98} 
M. L\"assig, 
J. Phys. Cond. Mat. \textbf{10} (1998), 9905.

\bibitem{blatter} 
G.\ Blatter, M.\ V.\ Feigel'man, V.\ B.\ Geshkenbein, A.\ I.\ Larkin, 
and V.\ M.\ Vinokur, 
Rev.\ Mod.\ Phys. \textbf{66} (1994), 1125. 

\bibitem{Cardy82} 
J.\ L.\ Cardy and S.\ Ostlund, 
Phys.\ Rev.\ B \textbf{25} (1982), 6899.

\bibitem{Toner90} 
J.\ Toner and D.\ P.\ DiVincenzo, 
Phys.\ Rev.\ B \textbf{41} (1990), 632.

\bibitem{review}
J.-Ph.\ Bouchaud, L.\ F.\ Cugliandolo, J.\ Kurchan, and M.\ M\'ezard in 
\textit{Spin Glasses and Random Fields}, ed.\ A.\ P.\ Young, 
(World Scientific, Singapore 1997); 
T.\ Giarmarchi and P.\ Le~Doussal, \textit{ibid.}

\bibitem{droplet} 
D.\ S.\ Fisher and D.\ A.\ Huse, 
Phys.\ Rev.\ B  \textbf{38} (1988), 386; 
for recent work see M.\ Palassini and A.\ P.\ Young, 
Phys.\ Rev.\  Lett.\ \textbf{83} (1999), 5126.

\bibitem{replica} 
M.\ M\'ezard, G.\ Parisi, and M.\ A.\ Virasoro, 
\textit{Spin Glass Theory and Beyond} 
(World Scientific, Singapore 1987); 
K.\ Binder and A.\ P.\ Young, 
Rev.\ Mod.\ Phys. \textbf{58} (1986), 801.

\bibitem{Lip86} 
R.\ Lipowsky and M.\ E.\ Fisher, 
Phys.\ Rev.\ Lett.\ \textbf{56} (1986), 472.

\bibitem{Wuttke91} 
J.\ Wuttke and R.\ Lipowsky, 
Phys.\ Rev.\ B \textbf{44} (1991), 13 042.

\bibitem{Dole91} 
S.\ Dietrich 
in \textit{Phase Transitions and Critical Phenomena}, 
edited by C.\ Domb and J.\ L.\ Lebowitz (Academic Press, San Diego 1988),
vol.\ 12; 
G.\ Forgacs, R.\ Lipowsky and Th.\ M.\ Nieuwenhuizen 
in \textit{Phase Transitions and Critical Phenomena}, 
edited by C.\ Domb and J.\ L.\ Lebowitz (Academic Press, San Diego 1991),
vol.\ 14.

\bibitem{new}
See for recent studies, e.g.,
S.\ Scheidl and Y.\ Din\c{c}er,
cond-mat/0006048;
P.\ Chauve, P.\ Le Doussal, and K.\ Wiese,
Phys.\ Rev.\ Lett.\ \textbf{86} (2001), 1785.

\bibitem{Balents} 
L.\ Balents and D.\ S.\ Fisher, 
Phys.\ Rev.\ B \textbf{48} (1993), 5949.

\bibitem{HuHe} 
D.\ A.\ Huse and C.\ L.\ Henley
Phys.\ Rev.\ Lett.\ \textbf{54} (1985), 2708.

\bibitem{Seppala98} 
E.\ T.\ Sepp\"al\"a,  V.\ Pet\"aj\"a, and M.\ J.\ Alava, 
Phys.\ Rev.\ E \textbf{58} (1998), R5217.

\bibitem{Fish91} 
D.\ S.\ Fisher and D.\ A.\ Huse, 
Phys.\ Rev.\ B \textbf{43} (1991), 10 728.

\bibitem{Hou1} 
For the spin glass case see 
J.\ Houdayer and O.\ C.\ Martin,  
Phys.\ Rev.\ Lett.\ \textbf{81} (1998), 2554.

\bibitem{Mez90} 
M.\ M\'ezard, 
J.\ Phys.\ France \textbf{51} (1990), 1831.

\bibitem{Hwa94} 
T.\ Hwa and D.\ S.\ Fisher, 
Phys.\ Rev.\ B \textbf{49} (1994), 3136.

\bibitem{Seppala00}
E.\ T.\ Sepp\"al\"a and M.\ J.\ Alava, 
Phys.\ Rev.\ Lett.\ \textbf{84} (2000), 3982.

\bibitem{unpublished} 
E.\ T.\ Sepp\"al\"a, M.\ J.\ Alava, and P.\ M.\ Duxbury, 
Phys.\ Rev.\ E, in press, e-print: cond-mat/0102318.

\bibitem{Parisi90} 
G.\ Parisi, 
J.\ Phys.\ France \textbf{51} (1990), 1595.

\bibitem{derrida} 
B.\ Derrida, 
Phys.\ Rev.\ B \textbf{24} (1981), 2613.

\bibitem{boumez} 
J.-P.\ Bouchaud and M. M\'ezard, 
J.\ Phys. A \textbf{30} (1997), 7997.

\bibitem{Gumbel} 
E.\ J.\ Gumbel, 
\textit{Statistics of Extremes}, 
(Columbia University Press, New York 1958). 

\bibitem{Shapir91} 
Y.\ Shapir, 
Phys.\ Rev.\ Lett.\ \textbf{66} (1991), 1473.

\bibitem{Joe98} 
P.\ J\"ogi and D.\ Sornette, 
Phys.\ Rev.\ E  \textbf{57} (1998), 6936.

\bibitem{Hua89} 
M.\ Huang, M.\ E.\ Fisher, and R.\ Lipowsky,
Phys.\ Rev.\ B \textbf{39} (1989), 2632.

\bibitem{Galambos} 
See also J.\ Galambos, 
\textit{The Asymptotic Theory of Extreme Order Statistics}, 
(John Wiley \& Sons, New York 1978), section II. 

\bibitem{DBL}
P.\ M.\ Duxbury, P.\ L.\ Leath, and P.\ D.\ Beale, 
Phys.\ Rev.\ B \textbf{36} (1987), 367; P.\ M.\ Duxbury, P.\ L.\ Leath, and 
P.\ D.\ Beale, Phys.\ Rev.\ Lett.\ \textbf{57} (1986), 1052.

\bibitem{KBM} 
J.\ A.\ Kim, M.\ A.\ Moore, and A.\ J.\ Bray,
Phys.\ Rev.\ A \textbf{44} (1991), 2345.

\bibitem{Rowlands}
S.\ C.\ Chapman, G.\ Rowlands, and N.\ W.\ Watkins,
cond-mat/0007275.

\bibitem{Bouchaud92} 
J.-Ph.\ Bouchaud and A.\ Georges, 
Phys.\ Rev.\ Lett.\ \textbf{68} (1992), 3908.

\bibitem{Alava96} 
M.\ J.\ Alava and P.\ M.\ Duxbury, 
Phys.\ Rev.\ B \textbf{54} (1996), 14990.

\bibitem{Raisanen98}
V.\  I.\  R\"ais\"anen, E.\ T.\ Sepp\"al\"a, M.\ J.\ Alava, 
and P.\ M.\ Duxbury, 
Phys.\ Rev.\ Lett.\ \textbf{80} (1998), 329. 

\bibitem{unpublished2} 
E.\ T.\ Sepp\"al\"a, M.\ J.\ Alava, and P.\ M.\ Duxbury, 
Phys.\ Rev.\ E \textbf{63} (2001), 036126. 

\bibitem{Alavaetal} 
M. Alava, P. Duxbury, C. Moukarzel, and H. Rieger, 
in \textit{Phase Transitions and Critical Phenomena}, 
edited by C.\ Domb and J.\ L.\ Lebowitz (Academic Press, San Diego 2001),
vol.\ 18.

\bibitem{network} 
J.\ C.\ Picard and H.\ D.\ Ratliff, 
Networks \textbf{5} (1975), 357. 

\bibitem{FordF} 
L.\ R.\ Ford and D.\ R.\ Fulkerson,
\textit{Flows in Networks},
(Princeton University Press, Princeton 1962)
see also any standard book on graph theory and flow problems.

\bibitem{Goltar88} 
A.\ V.\ Goldberg and R.\ E.\ Tarjan, 
J.\ Assoc.\ Comput.\ Mach. \textbf{35} (1988), 921.

\bibitem{resgraph}
Note that one can also use the level crossing effect to compute the
crossing points exactly, as noticed by Angl\`es d'Auriac and
Sourlas~\cite{Auriac}. This allows to use the residual graph of the
original groundstate for the further computations, and makes the
solution faster \cite{Alavaetal}.

\bibitem{Hansenetal}
S.\ Roux, A.\ Hansen, and E.\ Guyon, 
J.\ Phys. (Paris) \textbf{48} (1987), 2125.

\bibitem{Aizenman} 
M.\ Aizenman and J.\ Wehr, 
Phys.\ Rev.\ Lett.\ \textbf{62} (1989), 2503.

\bibitem{Barab}
A.-L.\ Barab{\'a}si and H.\ E.\ Stanley,
\textit{Fractal concepts in surface growth}, 
(Cambridge University Press, Cambridge, U.K. 1995).

\bibitem{spohn} 
See an example of effects of boundary conditions: 
M.\ Pr\"ahofer and H.\ Spohn,
Phys.\ Rev.\ Lett.\ \textbf{84} (2000), 4882.

\bibitem{Auriac}
J.\ C.\ Angl\`es d'Auriac and N.\ Sourlas,
Europhys.\ Lett. \textbf{39} (1997), 473.

\end{thebibliography}
\end{document}